\documentclass[journal,comsoc,draftcls,onecolumn,12pt,twoside]{IEEEtran}
\usepackage{multicol}
\usepackage{graphicx}
\usepackage{amssymb}
\usepackage{multirow}
\usepackage{tabularx}
\usepackage{acronym}
\usepackage{color}
\usepackage{lettrine}
\usepackage{mdwlist}
\usepackage{gensymb}
\usepackage{paralist}
\usepackage{cite}
\usepackage[cmex10]{amsmath}
\usepackage[font=footnotesize]{caption}
\usepackage[font=footnotesize]{subcaption}
\usepackage[utf8]{inputenc}
\usepackage{url}
\usepackage[utf8]{inputenc}
\usepackage[english]{babel}
\usepackage{amsthm}
\usepackage{hyperref}
\usepackage[ruled,vlined]{algorithm2e}

\usepackage[T1]{fontenc}

\ifCLASSINFOpdf
\else
\fi
\usepackage{amsmath}
\usepackage[cmintegrals]{newtxmath}
\begin{document}
%

\title{\color{blue} Spatial and Wavelength Division Joint Multiplexing System Design for Visible Light Communications}
%
%
%

\author{Cheng~Chen,~\IEEEmembership{Member,~IEEE,}
        Shenjie~Huang,~\IEEEmembership{Member,~IEEE,}
        Iman~Tavakkolnia,~\IEEEmembership{Member,~IEEE,}
        Majid~Safari,~\IEEEmembership{Senior Member,~IEEE,}
        and~Harald~Haas,~\IEEEmembership{Fellow,~IEEE}
\thanks{Cheng~Chen, Iman~Tavakkolnia and Harald~Haas are with the Department
of Electrical and Electrical Engineering, the University of Strathclyde, Glasgow, Scotland, UK.}
\thanks{Shenjie~Huang and Majid~Safari are with the institute for Digital Communications, School of Engineering, University of Edinburgh, Edinburgh, Scotland, UK.}
\thanks{Harald Haas acknowledges the financial support from the Wolfson Foundation, the Royal Society, and  Engineering and Physical Sciences Research Council (EPSRC) under Established Career Fellowship grant EP/R007101/1.}}

%
%

\markboth{}%
{}
%



\makeatletter
\newif\ifAC@uppercase@first%
\def\Aclp#1{\AC@uppercase@firsttrue\aclp{#1}\AC@uppercase@firstfalse}%
\def\AC@aclp#1{%
	\ifcsname fn@#1@PL\endcsname%
	\ifAC@uppercase@first%
	\expandafter\expandafter\expandafter\MakeUppercase\csname fn@#1@PL\endcsname%
	\else%
	\csname fn@#1@PL\endcsname%
	\fi%
	\else%
	\AC@acl{#1}s%
	\fi%
}%
\def\Acp#1{\AC@uppercase@firsttrue\acp{#1}\AC@uppercase@firstfalse}%
\def\AC@acp#1{%
	\ifcsname fn@#1@PL\endcsname%
	\ifAC@uppercase@first%
	\expandafter\expandafter\expandafter\MakeUppercase\csname fn@#1@PL\endcsname%
	\else%
	\csname fn@#1@PL\endcsname%
	\fi%
	\else%
	\AC@ac{#1}s%
	\fi%
}%
\def\Acfp#1{\AC@uppercase@firsttrue\acfp{#1}\AC@uppercase@firstfalse}%
\def\AC@acfp#1{%
	\ifcsname fn@#1@PL\endcsname%
	\ifAC@uppercase@first%
	\expandafter\expandafter\expandafter\MakeUppercase\csname fn@#1@PL\endcsname%
	\else%
	\csname fn@#1@PL\endcsname%
	\fi%
	\else%
	\AC@acf{#1}s%
	\fi%
}%
\def\Acsp#1{\AC@uppercase@firsttrue\acsp{#1}\AC@uppercase@firstfalse}%
\def\AC@acsp#1{%
	\ifcsname fn@#1@PL\endcsname%
	\ifAC@uppercase@first%
	\expandafter\expandafter\expandafter\MakeUppercase\csname fn@#1@PL\endcsname%
	\else%
	\csname fn@#1@PL\endcsname%
	\fi%
	\else%
	\AC@acs{#1}s%
	\fi%
}%
\edef\AC@uppercase@write{\string\ifAC@uppercase@first\string\expandafter\string\MakeUppercase\string\fi\space}%
\def\AC@acrodef#1[#2]#3{%
	\@bsphack%
	\protected@write\@auxout{}{%
		\string\newacro{#1}[#2]{\AC@uppercase@write #3}%
	}\@esphack%
}%
\def\Acl#1{\AC@uppercase@firsttrue\acl{#1}\AC@uppercase@firstfalse}
\def\Acf#1{\AC@uppercase@firsttrue\acf{#1}\AC@uppercase@firstfalse}
\def\Ac#1{\AC@uppercase@firsttrue\ac{#1}\AC@uppercase@firstfalse}
\def\Acs#1{\AC@uppercase@firsttrue\acs{#1}\AC@uppercase@firstfalse}
\robustify\Ac
\robustify\Aclp
\robustify\Acfp
\robustify\Acp
\robustify\Acsp
\robustify\Acl
\robustify\Acf
\robustify\Acs
\def\AC@@acro#1[#2]#3{%
	\ifAC@nolist%
	\else%
	\ifAC@printonlyused%
	\expandafter\ifx\csname acused@#1\endcsname\AC@used%
	\item[\protect\AC@hypertarget{#1}{\acsfont{#2}}] #3%
	\ifAC@withpage%
	\expandafter\ifx\csname r@acro:#1\endcsname\relax%
	\PackageInfo{acronym}{%
		Acronym #1 used in text but not spelled out in
		full in text}%
	\else%
	\dotfill\pageref{acro:#1}%
	\fi\\%
	\fi%
	\fi%
	\else%
	\item[\protect\AC@hypertarget{#1}{\acsfont{#2}}] #3%
	\fi%
	\fi%
	\begingroup
	\def\acroextra##1{}%
	\@bsphack
	\protected@write\@auxout{}%
	{\string\newacro{#1}[\string\AC@hyperlink{#1}{#2}]{\AC@uppercase@write #3}}%
	\@esphack
	\endgroup}
\makeatother

\acrodef{vlc}[VLC]{visible light communication}
\acrodef{lifi}[LiFi]{light fidelity}
\acrodef{dc}[DC]{direct current}
\acrodef{fft}[FFT]{fast Fourier transform}
\acrodef{ifft}[IFFT]{inverse fast Fourier transform}
\acrodef{dtft}[DTFT]{discrete-time Fourier transform}
\acrodef{dco}[DCO]{DC-biased optical}
\acrodef{ofdm}[OFDM]{orthogonal frequency division multiplexing}
\acrodef{led}[LED]{light-emitting diode}
\acrodef{pd}[PD]{photodiode}
\acrodef{csi}[CSI]{channel state information}
\acrodef{cp}[CP]{cyclic prefix}
\acrodef{mimo}[MIMO]{multiple-input multiple-output}
\acrodef{snr}[SNR]{signal-to-noise ratio}
\acrodef{e/o}[E/O]{electrical-to-optical}
\acrodef{apd}[APD]{avalanche photodiode}
\acrodef{qam}[QAM]{quadrature amplitude modulation}
\acrodef{isi}[ISI]{inter-symbol interference}
\acrodef{papr}[PAPR]{peak-to-average power ratio}
\acrodef{svd}[SVD]{singular-value decomposition}
\acrodef{smx}[SMX]{spatial multiplexing}
\acrodef{adr}[ADR]{angular diversity receiver}
\acrodef{wdm}[WDM]{wavelength division division}
\acrodef{rrc}[RRC]{root-square raised cosine}
\acrodef{los}[LoS]{line-of-sight}
\acrodef{nlos}[NLoS]{non-line-of-sight}
\acrodef{fov}[FoV]{field-of-view}
\acrodef{dof}[DoF]{degree-of-freedom}
\acrodef{2d}[2D]{two-dimensional}
\acrodef{3d}[3D]{three-dimensional}
\acrodef{ue}[UE]{user equipment}
\acrodef{swjm}[SWJM]{space \& wavelength division joint multiplexing}
\acrodef{bads}[BADS]{Bayesian adaptive direct search}
\acrodef{mads}[MADS]{mesh adaptive direct search}
\acrodef{scwd}[SCWD]{spatial clustering with wavelength division}
\acrodef{ber}[BER]{bit error rate}
\acrodef{kkt}[KKT]{Karush–Kuhn–Tucker}
\acrodef{rf}[RF]{radio frequency}
\acrodef{owc}[OWC]{optical wireless communication}
\acrodef{mhh}[MHH]{modified Hughes-Hartogs}
\acrodef{sd}[SD]{space division}
\acrodef{wd}[WD]{wavelength division}
\acrodef{pdf}[PDF]{probability density function}
\acrodef{pr}[PR]{pyramid receiver}
\acrodef{hr}[HR]{hemispheric receiver}
\acrodef{cie}[CIE]{International Commission on Illumination}
\acrodef{im/dd}[IM/DD]{intensity modulation with direct detection}
\acrodef{irs}[IRS]{intelligent reflecting surface}
\acrodef{ap}[AP]{access point}
\acrodef{iot}[IoT]{Internet-of-things}

\maketitle

\begin{abstract}
	The low-pass characteristics of front-end elements including light-emitting diodes (LEDs) and photodiodes (PDs) limit the transmission data rate of visible light communication (VLC) and Light Fidelity (LiFi) systems. Using multiplexing transmission techniques, such as spatial multiplexing (SMX) and wavelength division multiplexing (WDM), is a solution to overcome bandwidth limitation. {\color{blue} However, spatial correlation in optical wireless channels and optical filter bandpass shifts typically limit the achievable multiplexing gain in SMX and WDM systems, respectively. In this paper, we consider a multiple-input multiple output (MIMO) joint multiplexing VLC system that exploits available degrees-of-freedom (DoFs) across space, wavelength and frequency dimensions simultaneously. Instead of providing a new precoder/post-detector design, we investigate the considered joint multiplexing system from a system configuration perspective by tuning system parameters in both spatial and wavelength domains, such as LED positions and optical filter passband. We propose a novel spatial clustering with wavelength division (SCWD) strategy which enhances the MIMO channel condition. We propose to use a state-of-the-art black-box optimization tool: Bayesian adaptive direct search (BADS) to determine the desired system parameters, which can significantly improve the achievable rate. 
	The extensive numerical results demonstrate the superiority of the proposed method over conventional SMX and WDM VLC systems.}
\end{abstract}


%
\IEEEpeerreviewmaketitle

\section{Introduction}

\IEEEPARstart{W}{ith} the development of information technology, an increasing number of machine-type devices, wireless sensors and cloud services are deployed, which further increases the demand on wireless network capability \cite{9349624}. To meet the need of future wireless services, various new technologies for high speed wireless transmission have been proposed. \ac{vlc} and \ac{lifi} are among potential candidates \cite{Haas2016}. Apart from the use of the license-free optical spectrum and physical layer security feature, \ac{vlc} and \ac{lifi} can deliver multi-Gbps transmission data rate \cite{9489717}.
A major challenge of developing high performance \ac{vlc} and \ac{lifi} systems is the limited modulation bandwidth of \acp{led} and \acp{pd}. One of the solutions is to develop optical front-ends with a much wider bandwidth, such as GaN-based micro-\acp{led} \cite{9040548}. An alternative solution is to use \ac{smx} or \ac{wdm} techniques to transmit data with multiple parallel channels, which can significantly boost the aggregate data rate without expanding the modulation bandwidth.



{\color{blue} Several early studies on \ac{smx} with/without precoding/post-detection have been reported in \cite{6384613,7106482,6825137}. To include the frequency-domain characteristics, \ac{vlc} systems with \ac{mimo}-\ac{ofdm} have been investigated in \cite{7378876,7511379}. In addition, a coherent \ac{lifi} system with \ac{smx} has been considered in a recent study \cite{9408631}. 
Regarding research on \ac{wdm}-based \ac{vlc} systems, many successful experimental demonstrations were reported in the literature \cite{8675375,cossu20123}. An analytical work shows that the achievable data rate is limited by inter-channel crosstalk when a large number of wavelength divisions are used \cite{7728023}.  In order to mitigate the crosstalk, the usage of signal processing in \ac{mimo} systems have been proposed in a few studies \cite{8617505,lee2018deep}, where the \ac{wdm} channels are treated as a colour \ac{mimo} channel matrix and precoding/post-detection processes are used to diagonalise the multiplexing channel matrix. A recent study shows that it is possible to implement a \ac{wdm} system without using optical filters \cite{burton2021optical}.

\subsection{Related research}

Regarding the combination of \ac{smx} and \ac{wdm} \ac{vlc} systems, \ac{mimo}-\ac{vlc} systems with multi-colour \acp{led} have been considered to use \acp{dof} in both dimensions \cite{7862189,8742666}. These studies have proposed signal processing techniques such as optimal precoder designs under lighting constraints \cite{7862189} or a chromaticity-adaptive generalised spatial modulation scheme \cite{8742666}. In these studies, advanced signal processing techniques are generally designed to improve communication performance with a given set of system parameters.
Alternatively, the performance of \ac{vlc} systems can also be improved by changing system configurations, where key system parameters, such as \ac{led} position or optical filter passband, are carefully selected so that the probability of improved channel quality is increased. 
Several studies in multi-cell \ac{lifi} systems investigated the optimal system configurations in terms of \acp{ap} spatial deployment and \ac{led} parameters so that the system reliability, spectral efficiency or energy efficiency is maximised \cite{sharma2018optimal,niaz2016deployment,gismalla2021design,9130725}. Regarding the research on system configurations in \ac{vlc} \ac{mimo} systems, \acp{adr}, mirror diversity receiver and irregular \ac{pd} configurations are investigated to improve \ac{mimo} channel condition \cite{7109107,7794764,8904106}. System configurations of wavelength domain parameters in \ac{wdm} \ac{vlc} systems have been investigated in \cite{7728023,8546751}. Nevertheless, the above studies consider system configurations with only a few parameters in either the spatial or wavelength domain.

\subsection{Motivation and contributions}

Due to significant spatial correlation, the number of parallel channels in \ac{smx} \ac{vlc} systems is limited \cite{8904106}. On the other hand, it has been shown that the passband of a thin film  optical filter will shift to shorter wavelengths when the light incident angle is greater than 0\degree, which causes a severe wavelength mismatch in a \ac{wdm} \ac{vlc} system \cite{8546751}. By considering a spatial and wavelength division joint multiplexing \ac{vlc} system, the \ac{mimo} channel with severe spatial correlation can be decorrelated by the wavelength domain features. In addition, the excessive inter-colour interference in wavelength domain due to aforementioned passband shift issue can be mitigated by \ac{mimo} precoding and post-processing blocks. Consequently, the resultant number of parallel channels in joint multiplexing system can be increased and the corresponding achievable rates can be improved. 
Despite the performance improvement from the novel signal processing techniques, the combining features in spatial and wavelength domains have not been comprehensively investigated in \cite{7862189,8742666}. The research findings in \cite{sharma2018optimal,niaz2016deployment,gismalla2021design,9130725,7109107,7794764,8904106,8546751} also demonstrate the importance of system configuration in \ac{vlc}/\ac{lifi} systems, which has not been explored in a spatial and wavelength domain joint multiplexing \ac{vlc} system yet. Furthermore, it is complicated to design a \ac{vlc} multiplexing system using both spatial and wavelength domain features efficiently.
In this paper, a \ac{mimo}-\ac{ofdm} spatial and wavelength division joint multiplexing \ac{vlc} system is thoroughly studied from a system configuration perspective. To evaluate the impact of various parameters, a detailed framework of a \ac{mimo}-\ac{ofdm} joint multiplexing \ac{vlc} system is established considering the characteristics in the spatial, wavelength and frequency domains, which is unavailable in the literature. Based on the developed framework, the achievable rates with various system configurations are evaluated and compared. In particular, a unique \ac{scwd} configuration strategy is proposed which can achieve higher achievable rates compare to the other benchmark strategies. Furthermore, a \ac{bads} black-box optimisation tool has been used to search for system parameters that offer additional performance improvement. Compared to our previous study \cite{9120825}, a more detailed system model and more scenarios with practical concerns such as random user position/orientation are considered.
The contributions of this study are summarised as follows: 
\begin{itemize}
	\item A detailed framework is established for characterising \ac{vlc} \ac{mimo}-\ac{ofdm} joint multiplexing systems over space, wavelength and frequency domains. This framework allows researchers to evaluate the performance of a joint multiplexing system with a specific system configuration. 
	\item The system configurations of the considered \ac{vlc} \ac{mimo}-\ac{ofdm} joint multiplexing system are thoroughly investigated with random user position and device orientation. Both empirical parameter selections and parameter searching based on \ac{bads} algorithm are considered.
	\item Based on the idea of `division in either the spatial or wavelength domain', a \ac{scwd} strategy is proposed to efficiently use the \ac{dof} in both spatial and wavelength domains. The performance of the joint multiplexing systems using the \ac{scwd} strategy is compared with benchmark systems, which shows the superiority of the joint multiplexing system over \ac{smx} and \ac{wdm} techniques in terms of achievable rate. 
	\item  It has been found that systems with the \ac{scwd} strategy are superior to \ac{smx} and \ac{wdm} when two specific conditions are fulfilled: 1. When the achievable rates of \ac{smx} and \ac{wdm} are similar (with the same number of elements); 2. When the multiplexing gain improvement from excessively increasing numbers of \acp{led}/\acp{pd} is saturated for \ac{smx} and \ac{wdm} techniques. Insights into system configuration strategies and solutions are presented. 
\end{itemize}
}

The remainder of this paper is arranged as follows. Section~\ref{sec:MIMOOFDM_system_model} presents the considered \ac{mimo}-\ac{ofdm} system model. The \ac{mimo} channel model considering characteristics in space, wavelength and frequency domains are introduced in Section~\ref{sec:spatial_wavelength_channel}. The system configuration of the joint multiplexing system is thoroughly investigated in Section~\ref{sec:system_configurations}. 
The conclusions are drawn in Section~\ref{sec:conclusion}.

\section{MIMO-OFDM System model}
\label{sec:MIMOOFDM_system_model}

In this section, a conventional \ac{mimo}-\ac{ofdm} \ac{vlc} system model based on \ac{dco}-\ac{ofdm} is considered \cite{7378876}. A block diagram of the system is shown in Fig.~\ref{fig:mimo_ofdm_block_diagram}. Assuming there are $N_{\rm t}$ \acp{led} and $N_{\rm r}$ \acp{pd}, the maximum supported number of data streams will be  $I\leq\min(N_{\rm t},N_{\rm r})$. Considering a $K$-point \ac{fft} operation, the number of subcarriers carrying information bits is $\tilde{K}=K/2-1$. Firstly, a modulation block maps information bits to $M$-ary \ac{qam} symbols. Then, after a power control module and a precoding module, the signal vector on the $k$th subcarrier can be represented by:
\begin{align}
\mathbf{X}_k=\mathbf{F}_k\mathbf{Q}_k^{1/2}\mathbf{S}_k,~\text{for}~k=1,2,\cdots,\tilde{K},
\label{eq:X_mtx}
\end{align}
where $\mathbf{S}_k\in\mathbb{C}^{I\times 1}$ is the modulated symbol vector with unit variance; $\mathbf{Q}_k^{1/2}\in\mathbb{R}^{I\times I}$ is power control diagonal matrix and $\mathbf{F}_k\in\mathbb{C}^{N_{\rm t}\times I}$ is a precoding matrix, which must be a unity matrix. Note that the \ac{qam} symbol on the $k$th subcarrier of the $i$th data stream is defined as ${S}_{i}[k]$, which is also the $i$th element of $\mathbf{S}_k$. Thus, the signal value on the $k$th subcarrier for the $n_{\rm t}$th \ac{led} can be written as:
\begin{align}
{X}_{n_{\rm t}}[k]=\sum_{i=1}^{I}{F}_{n_{\rm t},i}[k]\sqrt{q_{i}[k]}{S}_{i}[k],
\label{eq:transmit_signal1}
\end{align}
where  $\sqrt{q_{i}[k]}$ is the $(i,i)$-th entry of $\mathbf{Q}_k^{1/2}$ and ${F}_{n_{\rm t},i}[k]$ is the $(n_{\rm t},i)$-th entry of $\mathbf{F}_k$. To guarantee a real-value time-domain signal, the Hermitian symmetry condition must be fulfilled, which requires: $X_{n_{\rm t}}[0]=X_{n_{\rm t}}[K/2]=0$ and $X_{n_{\rm t}}[k]=X_{n_{\rm t}}^{*}[K-k]$ for $k=K/2+1,K/2+2,\cdots,K-1$. Next, the frequency-domain signal is converted to time-domain for transmission by using an \ac{ifft} operation:
\begin{align}
x_{n_{\rm t}}[n]=\frac{1}{\sqrt{K}}\sum_{k=0}^{K-1}{X}_{n_{\rm t}}[k]\mathrm{e}^{\frac{2\pi nkj}{K}},
\label{eq:transmit_signal2}
\end{align}
where $j=\sqrt{-1}$ is the imaginary number. Due to the limited dynamic range of each \ac{led}, we normalise and constrain the signal variance not to be greater than unity: $\mathbb{E}\left\{x^2_{n_{\rm t}}[n]\right\}\leq 1$, where $\mathbb{E}\left\{\cdot\right\}$ refers to the expectation operator. Note that the optical power constraint is considered in Section~\ref{subsec:e_to_o_conversion}. 
To avoid inter-frame interference and \ac{isi}, a \ac{cp} is added to the beginning of each time-domain \ac{ofdm} frame. After the precoding and \ac{ifft} operations, a clipping operation is enforced to limit the signal \ac{papr}:
\begin{figure*}[!t]
	\begin{center}
		\includegraphics[width=0.99\textwidth]{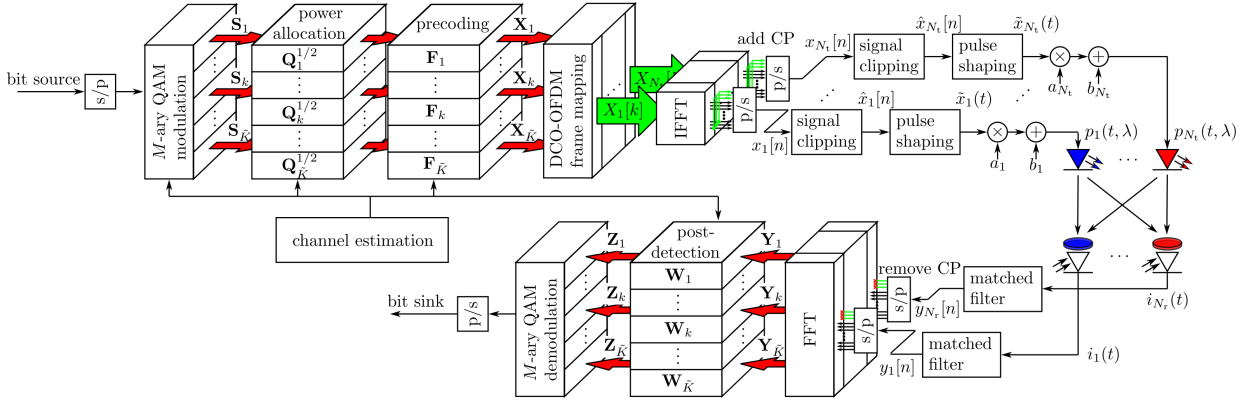}
	\end{center}
	\caption{Block diagram of a MIMO-OFDM VLC system based on DCO-OFDM.}
	\label{fig:mimo_ofdm_block_diagram}
	\vspace{-7mm}
\end{figure*}
\begin{align}
\hat{x}_{n_{\rm t}}[n]=\left\{\begin{array}{lr} \kappa_{\rm t} &: x_{n_{\rm t}}[n]\geq \kappa_{\rm t} \\
x_{n_{\rm t}}[n] &: \kappa_{\rm b}<x_{n_{\rm t}}[n]<\kappa_{\rm t} \\
\kappa_{\rm b} &:x_{n_{\rm t}}[n]\leq \kappa_{\rm b}\end{array}
\right.,
\label{eq:clipping1}
\end{align}
where $\kappa_{\rm t}$ and $\kappa_{\rm b}$ are the top and bottom clipping levels. According to the Bussgang theorem, the non-linear clipping operation can be approximated by:
\begin{align}
\hat{x}_{n_{\rm t}}[n]=\eta x_{n_{\rm t}}[n]+\mathfrak{n}_{n_{\rm t}}^{\rm clip}[n],
\label{eq:clipping2}
\end{align}
where $\eta$ is the clipping attenuation factor and $\mathfrak{n}_{n_{\rm t}}^{\rm clip}[n]$ is the clipping noise at the $n_{\rm t}$th \ac{led} \cite{tsh1302} which follows a normal distribution with a zero mean and a variance of $\sigma_{\rm clip}^2$. The value of $\mathfrak{n}_{n_{\rm t}}^{\rm clip}[n]$ and $\sigma_{\rm clip}^2$ can be calculated analytically \cite{Dimitrov2012}. Then, the clipped electric signal is converted to an optical signal and launched to the optical wireless channel. At the receiver side, a fraction of the optical signals are detected by the \acp{pd}. The detected signal by the $n_{\rm r}$th \ac{pd} can be calculated as:
\begin{align}
y_{n_{\rm r}}[n]=\mathfrak{n}^{\rm rx}_{n_{\rm r}}[n]+\sum_{n_{\rm t}=1}^{N_{\rm t}}\hat{x}_{n_{\rm t}}[n]\otimes h_{n_{\rm r},n_{\rm t}}[n],
\end{align}
where $\mathfrak{n}^{\rm rx}_{n_{\rm r}}[n]$ is the receiver noise, $h_{n_{\rm r},n_{\rm t}}[n]$ is the discrete channel impulse response between the $n_{\rm r}$th \ac{pd} and $n_{\rm t}$th \ac{led} and $\otimes$ refers to the convolution operator. The receiver noise follows a normal distribution with a zero mean and a variance of ${\sigma}_{{\rm rx},n_{\rm r}}^2$. The considered receiver noise is composed of the shot noise and thermal noise. After the reception of the signal, the \ac{cp} is removed. The time-domain signal is converted back to the frequency-domain using the \ac{fft} operation as:
\begin{align}
{Y}_{n_{\rm r}}[k]=\frac{1}{\sqrt{K}}\sum_{k=0}^{K-1}y_{n_{\rm r}}[n]\mathrm{e}^{-\frac{2\pi nkj}{K}}.
\end{align}
{\color{blue} The addition of \acp{cp} leads to the circular convolution relationship between the time-domain signal and the channel impulse response. This circular convolution can be converted to a multiplication relationship in the frequency-domain.} Therefore, the conversion between ${Y}_{n_{\rm r}}[k]$ and ${X}_{n_{\rm t}}[k]$ can also be evaluated in the frequency-domain directly. The received signal vector on the $k$th subcarrier $\mathbf{Y}_k\in\mathbb{C}^{N_{\rm r}\times 1}$ can be calculated by:
\begin{align}
\mathbf{Y}_k=\mathbf{H}_k\left(\eta\mathbf{X}_k+\mathbf{E}_k\right)+\mathbf{N}_k,
\label{eq:Y_mtx}
\end{align}
where $\mathbf{H}_k\in\mathbb{C}^{N_{\rm r}\times N_{\rm t}}$ is the frequency-domain channel matrix, $\mathbf{E}_k\in\mathbb{C}^{N_{\rm t}\times 1}$ is the frequency-domain clipping noise vector and  $\mathbf{N}_k\in\mathbb{C}^{N_{\rm r}\times 1}$ is the frequency-domain receiver noise vector. The $(n_{\rm r},n_{\rm t})$-th entry of $\mathbf{H}_k$ is denoted as $H_{n_{\rm r},n_{\rm t}}[k]$, the $n_{\rm t}$th entry of $\mathbf{E}_k$ is denoted as $N_{n_{\rm t}}^{\rm clip}[k]$ and the $n_{\rm r}$th entry of $\mathbf{N}_k$ is denoted as $N_{n_{\rm r}}^{\rm rx}[k]$. In addition, $H_{n_{\rm r},n_{\rm t}}[k]$, $N_{n_{\rm t}}^{\rm clip}[k]$ and $N_{n_{\rm r}}^{\rm rx}[k]$ are \ac{fft} of $h_{n_{\rm r},n_{\rm t}}[n]$, $\mathfrak{n}^{\rm clip}_{n_{\rm t}}[n]$ and $\mathfrak{n}^{\rm rx}_{n_{\rm r}}[n]$, respectively. Finally, a \ac{mimo} post-detection matrix $\mathbf{W}_k\in\mathbb{C}^{I\times N_{\rm r}}$ is used to retrieve the transmitted data symbol vectors:
\begin{align}
\mathbf{Z}_k&=\mathbf{W}_k\mathbf{Y}_k=\eta\mathbf{W}_k\mathbf{H}_k\mathbf{F}_k\mathbf{Q}_k^{1/2}\mathbf{S}_k+\mathbf{W}_k\mathbf{H}_k\mathbf{E}_k+\mathbf{W}_k\mathbf{N}_k.
\end{align}
Thus, the data symbol of the $i$th data stream on the $k$th subcarrier can be written as:
\begin{align}
&Z_i[k]=\eta\sum_{n_{\rm r}=1}^{N_{\rm r}}\sum_{n_{\rm t}=1}^{N_{\rm t}} {W}_{i,n_{\rm r}}[k]H_{n_{\rm r},n_{\rm t}}[k]{F}_{n_{\rm t},i}[k]\sqrt{q_{i}[k]}S_{i}[k]+\sum_{n_{\rm r}=1}^{N_{\rm r}}\sum_{n_{\rm t}=1}^{N_{\rm t}}{W}_{i,n_{\rm r}}[k]H_{n_{\rm r},n_{\rm t}}[k]{N}^{\rm clip}_{n_{\rm t}}[k] \nonumber \\ &+\eta\sum_{n_{\rm r}=1}^{N_{\rm r}}\sum_{n_{\rm t}=1}^{N_{\rm t}} \sum_{\hat{i}=1,\hat{i}\neq i}^{I}{W}_{i,n_{\rm r}}[k]H_{n_{\rm r},n_{\rm t}}[k]{F}_{n_{\rm t},\hat{i}}[k]\sqrt{q_{\hat{i}}[k]}S_{\hat{i}}[k]+\sum_{n_{\rm r}=1}^{N_{\rm r}}{W}_{i,n_{\rm r}}[k]{N}^{\rm rx}_{n_{\rm r}}[k],
\end{align}
where ${W}_{i,n_{\rm r}}[k]$ is the $(i,n_{\rm r})$-th entry of $\mathbf{W}_k$, the first term on the right-hand side of the equality is the desired signal, the second term is the equivalent clipping noise, the third term is the interference from other multiplexing channels and the last term corresponds to the equivalent receiver noise. Therefore, the corresponding \ac{snr} of the $i$th data stream on the $k$th subcarrier can be calculated by:
	\begin{align}
	\gamma_i[k]&=\left(\eta^2\left|\sum_{n_{\rm r}=1}^{N_{\rm r}}\sum_{n_{\rm t}=1}^{N_{\rm t}} {W}_{i,n_{\rm r}}[k]H_{n_{\rm r},n_{\rm t}}[k]{F}_{n_{\rm t},i}[k]\right|^2q_{i}[k]\right)\left(\sigma_{\rm clip}^2\sum\limits_{n_{\rm r}=1}^{N_{\rm r}}\sum\limits_{n_{\rm t}=1}^{N_{\rm t}}\left|{W}_{i,n_{\rm r}}[k]H_{n_{\rm r},n_{\rm t}}[k]\right|^2\right. \nonumber\\ 
	&\left.+\eta^2\sum\limits_{\hat{i}=1,\hat{i}\neq i}^{I}\left|\sum\limits_{n_{\rm r}=1}^{N_{\rm r}}\sum\limits_{n_{\rm t}=1}^{N_{\rm t}} {W}_{i,n_{\rm r}}[k]H_{n_{\rm r},n_{\rm t}}[k]{F}_{n_{\rm t},\hat{i}}[k]\right|^2q_{\hat{i}}[k]+\sum\limits_{n_{\rm r}=1}^{I}\left|{W}_{i,n_{\rm r}}[k]\right|^2{\sigma}_{{\rm rx},n_{\rm r}}^2\right)^{-1},
	\label{eq:SNR}
	\end{align}
{\color{blue} In this study, the well-known \ac{svd}-based precoding and post-detection are used, which remove inter-channel interference and convert \ac{mimo} channels to orthogonal parallel channels.}


\section{Spatial and wavelength characteristics of MIMO-OFDM channel}
\label{sec:spatial_wavelength_channel}
In this section, we introduce the characteristics of the \ac{mimo}-\ac{ofdm} channel $\mathbf{H}_k$ introduced in Section~\ref{sec:MIMOOFDM_system_model}. The relationship between the channel and parameters in the space, wavelength and frequency domains is considered. Firstly, we consider the discrete samples forwarded to the $n_{\rm t}$th \ac{led} and its driving circuit. The discrete samples $\hat{x}_{n_{\rm t}}[n]$ are converted to a continuous analogue signal via a pulse shaping process:
\begin{align}
\tilde{x}_{n_{\rm t}}(t)=\sum_{n=-\infty}^{\infty}\hat{x}_{n_{\rm t}}[n]g(t-nT_{\rm s}),
\label{eq:x_continuous}
\end{align}
where $g(t)$ is the impulse response of the signal pulse and $T_{\rm s}$ is the symbol period. Before feeding the analogue signal, $\tilde{x}_{n_{\rm t}}(t)$ is amplified by a factor of $a_{n_{\rm t}}$ and a \ac{dc}-bias of $b_{n_{\rm t}}$ is added in the driving circuit. The optical signal of the $n_{\rm t}$th \ac{led} can be written as:
\begin{align}
p_{n_{\rm t}}(t,\lambda)=\mathcal{S}_{n_{\rm t}}^{\rm led}(\lambda)\left(a_{n_{\rm t}}\tilde{x}_{n_{\rm t}}(t)+b_{n_{\rm t}}\right)\otimes h^{\rm led}(t),
\label{eq:p_t}
\end{align}
where $\mathcal{S}_{n_{\rm t}}^{\rm led}(\lambda)$ is the normalized spectral density of the $n_{\rm t}$th \ac{led} at wavelength $\lambda$ and $h^{\rm led}(t)$ is the impulse response of the \ac{led}. After the emission of the optical signal to the wireless channel, a fraction of the signal is detected by the \acp{pd} on the receiver side. The output photocurrent of the $n_{\rm r}$th \ac{pd} can be written as:
\begin{align}
i_{n_{\rm r}}(t)&=\int\limits_{\lambda_{\rm min}}^{\lambda_{\rm max}}\mathcal{R}^{\rm pd}(\lambda)h^{\rm pd}(t)\otimes {h}_{n_{\rm r},n_{\rm t}}^{\rm ow}(t,\lambda)\otimes
p_{n_{\rm t}}(t,\lambda) \mathrm{d}\lambda,
\label{eq:i_t}
\end{align}
where $\mathcal{R}^{\rm pd}(\lambda)$ is the spectral \ac{pd} responsivity, $h^{\rm pd}(t)$ is the low-pass impulse response of the \ac{pd} and ${h}_{n_{\rm r},n_{\rm t}}^{\rm ow}(t,\lambda)$ is the optical wireless channel impulse response. Since the optical transmission operates in a wide spectrum region, the final photocurrent is the result of an integration over the involved spectrum region. Then the photocurrent is forwarded to a matched filter, where the detected waveform is convolved with the signal pulse $g(t)$ and the discrete signal is obtained by sampling at $nT_{\rm s}$. By applying a discrete-time unit impulse function input $\hat{x}_{n_{\rm t}}[n]=\delta[n]$ to \eqref{eq:x_continuous} and inserting \eqref{eq:x_continuous}, \eqref{eq:p_t} into \eqref{eq:i_t}, the continuous channel impulse response before sampling can be calculated as:
\begin{align}
&h_{n_{\rm r},n_{\rm t}}(t)=\int\limits_{\lambda_{\rm min}}^{\lambda_{\rm max}}\mathcal{R}^{\rm pd}(\lambda)\mathcal{S}_{n_{\rm t}}^{\rm led}(\lambda)g(t)\otimes h^{\rm pd}(t) \otimes {h}_{n_{\rm r},n_{\rm t}}^{\rm ow}(t,\lambda) \otimes h^{\rm led}(t) \otimes \left(a_{n_{\rm t}}g(t)+b_{n_{\rm t}}\right)\mathrm{d}\lambda. 
\label{eq:h_t}
\end{align}
Noting that the convolution operation is with respect to $t$, but the integral is with respect to $\lambda$. The Fourier transform of \eqref{eq:h_t} can be calculated as:
\begin{align}
&H_{n_{\rm r},n_{\rm t}}(f)=\int\limits_{\lambda_{\rm min}}^{\lambda_{\rm max}}\mathcal{R}^{\rm pd}(\lambda)\mathcal{S}_{n_{\rm t}}^{\rm led}(\lambda)G(f)H^{\rm pd}(f) {H}_{n_{\rm r},n_{\rm t}}^{\rm ow}(f,\lambda)  H^{\rm led}(f) \left(a_{n_{\rm t}}G(f)+b_{n_{\rm t}}\delta(f)\right)\mathrm{d}\lambda,
\label{eq:H_f}
\end{align}
where $G(f)$, $H^{\rm pd}(f)$, ${H}_{n_{\rm r},n_{\rm t}}^{\rm ow}(f,\lambda)$ and $H^{\rm led}(f)$ are the \ac{fft} of $g(t)$, $h^{\rm pd}(t)$, ${h}_{n_{\rm r},n_{\rm t}}^{\rm ow}(t,\lambda)$ and $h^{\rm led}(t)$, respectively. Based on the definition of \ac{dtft}, the channel transfer function after the matched filter sampling at $nT_{\rm s}$ is a periodic summation of $H_{n_{\rm r},n_{\rm t}}(f)$ with a period of $1/T_{\rm s}$ as: $H_{n_{\rm r},n_{\rm t}}^{1/T_{\rm s}}(f)=\sum_{l=-\infty}^{\infty}H_{n_{\rm r},n_{\rm t}}(f-l/T_{\rm s})$. Thus the channel transfer function between the $n_{\rm t}$th {LED} and the $n_{\rm r}$th PD on the $k$th subcarrier can be calculated as:
\begin{align}
{H}_{n_{\rm r},n_{\rm t}}[k]&=H_{n_{\rm r},n_{\rm t}}^{1/T_{\rm s}}\left(\frac{k}{KT_{\rm s}}\right)=\sum_{l=-\infty}^{\infty}H_{n_{\rm r},n_{\rm t}}\left(\frac{k}{KT_{\rm s}}-\frac{l}{T_{\rm s}}\right),
\label{eq:H_nr_nt_k}
\end{align}
for $k=0,1,\cdots,K-1$. {\color{blue} Although the limits of the summation in \eqref{eq:H_nr_nt_k} are from $-\infty$ to $\infty$, the band-limited signal pulse $G(f)$ makes most of the terms in the summation equal zero. For example, the used \ac{rrc} pulse in this work makes the shifted channel frequency response \eqref{eq:H_f} equals zero for $f\in\left(-\infty,-\frac{\alpha+1}{2T_{\rm s}}+\frac{l}{T_{\rm s}}\right)\cup\left(\frac{\alpha+1}{2T_{\rm s}}+\frac{l}{T_{\rm s}},\infty\right)$. The subcarrier index of the transfer function falls in the region of $[0,K-1]$, which is within the frequency range of $[0,1/T_{\rm s})$. This implies that only the terms with $l=0$ and $1$ are non-zero in \eqref{eq:H_nr_nt_k}.}


\begin{figure}[!t]
	\centering
	\begin{subfigure}{.43\textwidth}
		\centering
		\includegraphics[width=.9\linewidth]{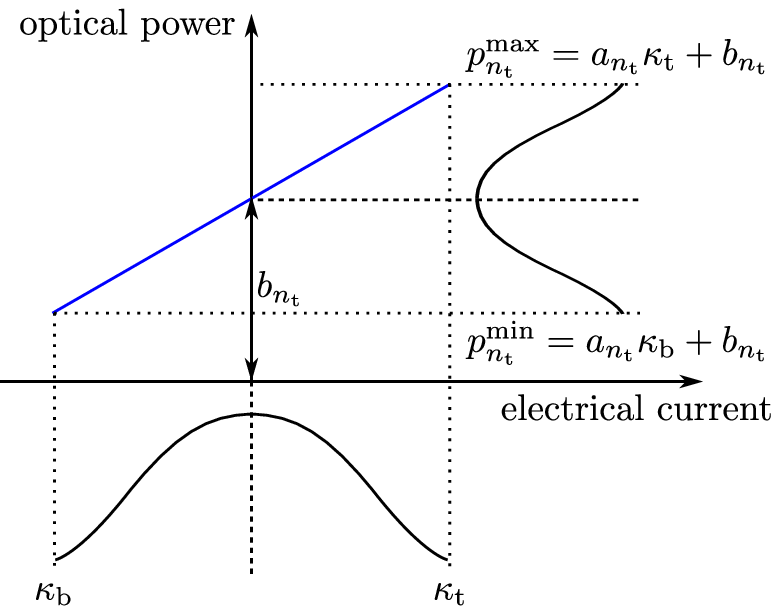}
		\caption{}
		\label{fig:elec_to_optical}
	\end{subfigure}%
	\begin{subfigure}{.57\textwidth}
		\centering
		\includegraphics[width=.9\linewidth]{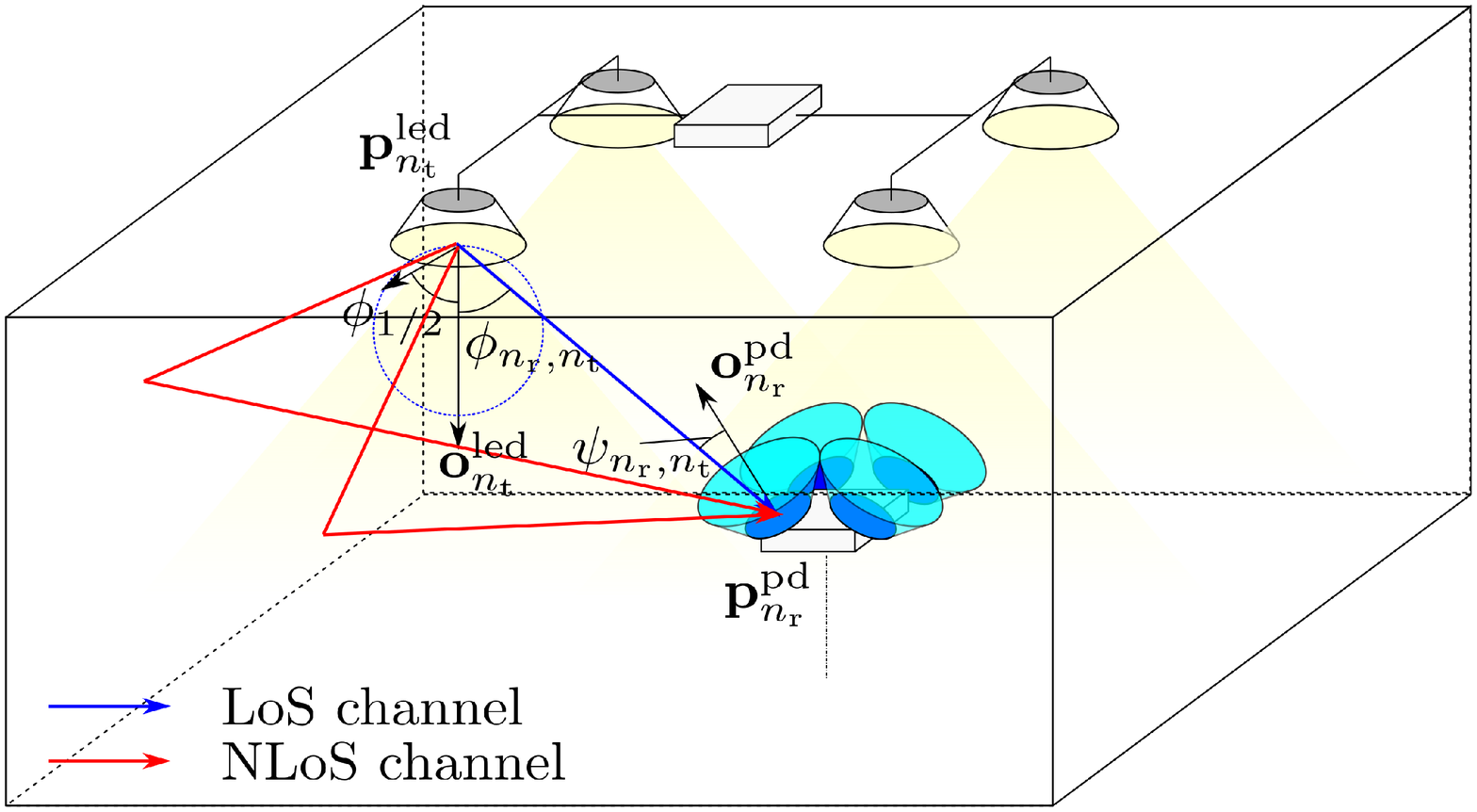}
		\caption{}
		\label{fig:spatial_geo}
	\end{subfigure}
	\caption{(a) Electrical to optical signal conversion. (b) Spatial-domain characteristics with LoS and NLoS channel components.}
	\label{fig:elec_to_optical_spatial_geo}
	\vspace{-7mm}
\end{figure} 
\subsection{LED electrical-to-optical conversion}
\label{subsec:e_to_o_conversion}
In \eqref{eq:p_t}, the choice of $a_{n_{\rm t}}$ and $b_{n_{\rm t}}$ should map the peak values of the input signal to the peak values of the optical output to maximise the signal power, as illustrated in Fig.~\ref{fig:elec_to_optical}. Assuming the $n_{\rm t}$th \ac{led} has a maximum and minimum optical levels of $p_{n_{\rm t}}^{\rm max}$ and $p_{n_{\rm t}}^{\rm min}$, the following mapping equations can be used: $a_{n_{\rm t}}\kappa_{\rm b}+b_{n_{\rm t}}=p_{n_{\rm t}}^{\rm min},~a_{n_{\rm t}}\kappa_{\rm t}+b_{n_{\rm t}}=p_{n_{\rm t}}^{\rm max}$,
which lead to:
\begin{align}
 a_{n_{\rm t}}=\frac{p_{n_{\rm t}}^{\rm max}-p_{n_{\rm t}}^{\rm min}}{\kappa_{\rm t}-\kappa_{\rm b}}, ~
 b_{n_{\rm t}}=\frac{p_{n_{\rm t}}^{\rm min}\kappa_{\rm t}-p_{n_{\rm t}}^{\rm max}\kappa_{\rm b}}{\kappa_{\rm t}-\kappa_{\rm b}}.
 \label{eq:e_to_o_conversion}
\end{align}
By using \eqref{eq:p_t}, the average optical power of the $n_{\rm t}$th \ac{led} can be calculated by:
\begin{align}
\bar{p}_{n_{\rm t}}=\lim_{T\to\infty}\int\limits_{-T}^{T}\int\limits_{\lambda_{\rm min}}^{\lambda_{\rm max}}\frac{p_{n_{\rm t}}(t,\lambda)}{2T}\mathrm{d}\lambda \mathrm{d}t=a_{n_{\rm t}}\mathbb{E}\left\{\hat{x}_{n_{\rm t}}[n]\right\}+b_{n_{\rm t}},
\label{eq:average_optical_power}
\end{align}
where $\mathbb{E}\left\{\hat{x}_{n_{\rm t}}[n]\right\}$ is the expectation of the clipped signal, which can be evaluated analytically \cite{Dimitrov2012}. For simplicity of the analysis, we consider a symmetric clipping ($\kappa_{\rm t}=\kappa$ and $\kappa_{\rm b}=-\kappa$) and a zero minimum optical power level ($p_{n_{\rm t}}^{\rm min}=0$). This leads to simplification to \eqref{eq:e_to_o_conversion} and \eqref{eq:average_optical_power} as $a_{n_{\rm t}}=p_{n_{\rm t}}^{\rm max}/2\kappa$, $b_{n_{\rm t}}=p_{n_{\rm t}}^{\rm max}/2=\bar{p}_{n_{\rm t}}$ and $\bar{p}_{n_{\rm t}}=b_{n_{\rm t}}$. If we consider $\bar{p}_{n_{\rm t}}$ as the given parameter, the scaling factor and \ac{dc}-bias can be calculated by $a_{n_{\rm t}}=\bar{p}_{n_{\rm t}}/\kappa$ and $\bar{p}_{n_{\rm t}}$.

\subsection{Spatial-domain characteristics}
\label{subsec:spatial_characteristics}
 The spatial-domain characteristics are primarily determined by the optical wireless channel between the \acp{led} and the \acp{pd}, which can be decomposed into a \ac{los} component and \ac{nlos} components, as shown in Fig.~\ref{fig:spatial_geo}. The corresponding transfer function can be defined as: ${H}_{n_{\rm r},n_{\rm t}}^{\rm ow}(f,\lambda)={H}_{n_{\rm r},n_{\rm t}}^{\rm LoS}(f,\lambda)+{H}_{n_{\rm r},n_{\rm t}}^{\rm NLoS}(f,\lambda)$. The \ac{los} corresponds to the signal propagation directly from the \acp{led} to the \acp{pd}, which dominates the \ac{mimo} channel in most cases. The frequency response can be calculated as \cite{s1601}:
 \begin{align}
  {H}_{n_{\rm r},n_{\rm t}}^{\rm LoS}(f,\lambda)=\frac{\left(m_{\rm led}+1\right)A_{\rm pd}\mathbf{1}_{\rm v}}{2\pi D_{n_{\rm r},n_{\rm t}}^2}\exp(-j 2\pi f\tau_{n_{\rm r},n_{\rm t}})\mathcal{G}_{n_{\rm r}}^{\rm of}(\lambda,\psi_{n_{\rm r},n_{\rm t}})\cos^{m_{\rm led}}\phi_{n_{\rm r},n_{\rm t}}\cos^{m_{\rm fov}}\psi_{n_{\rm r},n_{\rm t}},
  \label{eq:LoS_channel}
 \end{align}
where $A_{\rm pd}$ is the active area of the \ac{pd}, $m_{\rm led}$ is the Lambertian emission order of the \ac{led}, $m_{\rm fov}$ is the \ac{fov} coefficient \cite{7109107}, $\mathbf{1}_{\rm v}$ is a visibility function, $\mathcal{G}_{n_{\rm r}}^{\rm of}(\lambda,\psi)$ is the transmittance of the optical filter mounted on the $n_{\rm r}$th \ac{pd} and $D_{n_{\rm r},n_{\rm t}}$, $\phi_{n_{\rm r},n_{\rm t}}$, $\psi_{n_{\rm r},n_{\rm t}}$, $\tau_{n_{\rm r},n_{\rm t}}$ are the Euclidean distance, radiant angle, incident angle, time delay between the $n_{\rm t}$th \ac{led} and the $n_{\rm r}$th \ac{pd}, respectively. The Lambertian emission order $m_{\rm led}$ is related to the \ac{led} half-power seminangle by $m_{\rm led}=-1/\log_2(\cos(\phi_{1/2}))$. The time delay can be calculated by $\tau_{n_{\rm r},n_{\rm t}}=D_{n_{\rm r},n_{\rm t}}/c$, where $c=3\times 10^8$~m/s is the speed of light. The visibility function is defined as:
\begin{align}
 \mathbf{1}_{\rm v}=\left\{\begin{array}{lr} 1 &: \phi<\pi/2~\text{and}~\psi<\pi/2 \\
 0 &: \text{otherwise} \end{array}
 \right.,
\end{align}
which forces the channel to be zero when either $\phi$ or $\psi$ exceed $pi/2$. The value of \eqref{eq:LoS_channel} is directly determined by the positions and orientations of the \acp{led} and \acp{pd}. The trigonometric and distance terms in \eqref{eq:LoS_channel} can be calculated by \cite{bkklm9301}:
\begin{align}
&D_{n_{\rm r},n_{\rm t}}=\left|\left|\mathbf{p}_{n_{\rm t}}^{\rm led}-\mathbf{p}_{n_{\rm r}}^{\rm pd}\right|\right|,~
\cos\phi_{n_{\rm r},n_{\rm t}}=\frac{\mathbf{o}_{n_{\rm t}}^{\rm led}}{D_{n_{\rm r},n_{\rm t}}}\cdot (\mathbf{p}_{n_{\rm r}}^{\rm pd}-\mathbf{p}_{n_{\rm t}}^{\rm led}),~\cos\psi_{n_{\rm r},n_{\rm t}}=\frac{\mathbf{o}_{n_{\rm r}}^{\rm pd}}{D_{n_{\rm r},n_{\rm t}}}\cdot (\mathbf{p}_{n_{\rm t}}^{\rm led}-\mathbf{p}_{n_{\rm r}}^{\rm pd}),
\end{align}
where $\mathbf{p}_{n_{\rm t}}^{\rm led}$ and $\mathbf{o}_{n_{\rm t}}^{\rm led}$ are the position and orientation vectors of the $n_{\rm t}$th \ac{led}, respectively; $\mathbf{p}_{n_{\rm r}}^{\rm pd}$ and $\mathbf{o}_{n_{\rm r}}^{\rm pd}$ are the position and orientation vectors of the $n_{\rm r}$th \ac{pd}, respectively; $\{\cdot\}$ refers to the vector dot product and $||\cdot||$ refers to the Euclidean norm. Note that the characteristics of $\mathcal{G}_{n_{\rm r}}^{\rm of}(\lambda,\psi)$ are affected by features in both the spatial and wavelength domains, which will be covered in Section~\ref{subsec:wavelength_domain_characteristics}. {\color{blue} The \ac{nlos} channel responses correspond to the signal propagation via reflections by the room internal surfaces, which can be evaluated using an efficient frequency-domain simulation method \cite{s1601}. }


\subsection{Wavelength-domain characteristics}
\label{subsec:wavelength_domain_characteristics}

\begin{figure}
	\centering
	\begin{subfigure}{.55\textwidth}
		\centering
		\includegraphics[width=.9\linewidth]{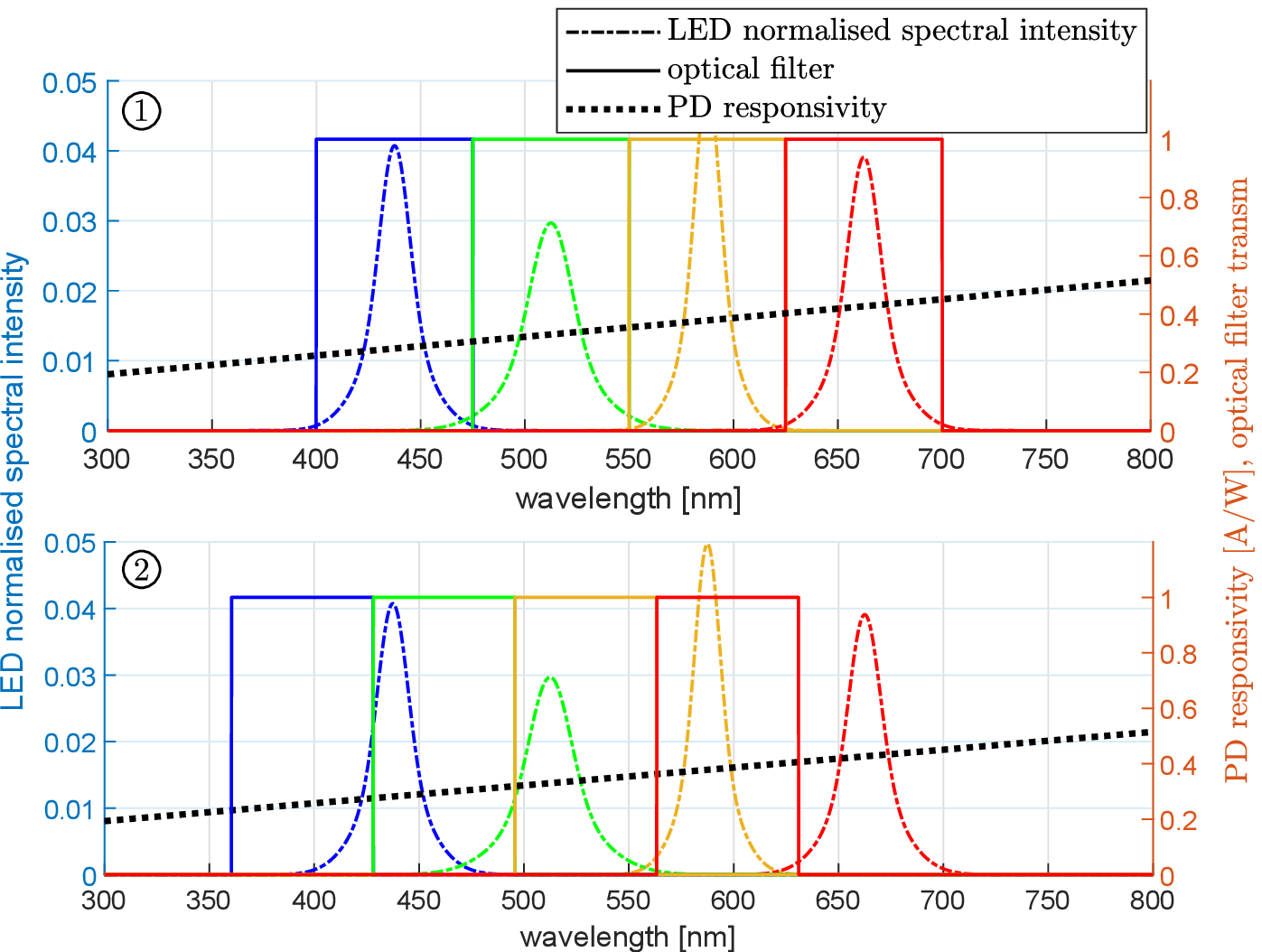}
		\caption{}
		\label{fig:wavelength_domain}
	\end{subfigure}%
	\begin{subfigure}{.45\textwidth}
		\centering
		\includegraphics[width=.9\linewidth]{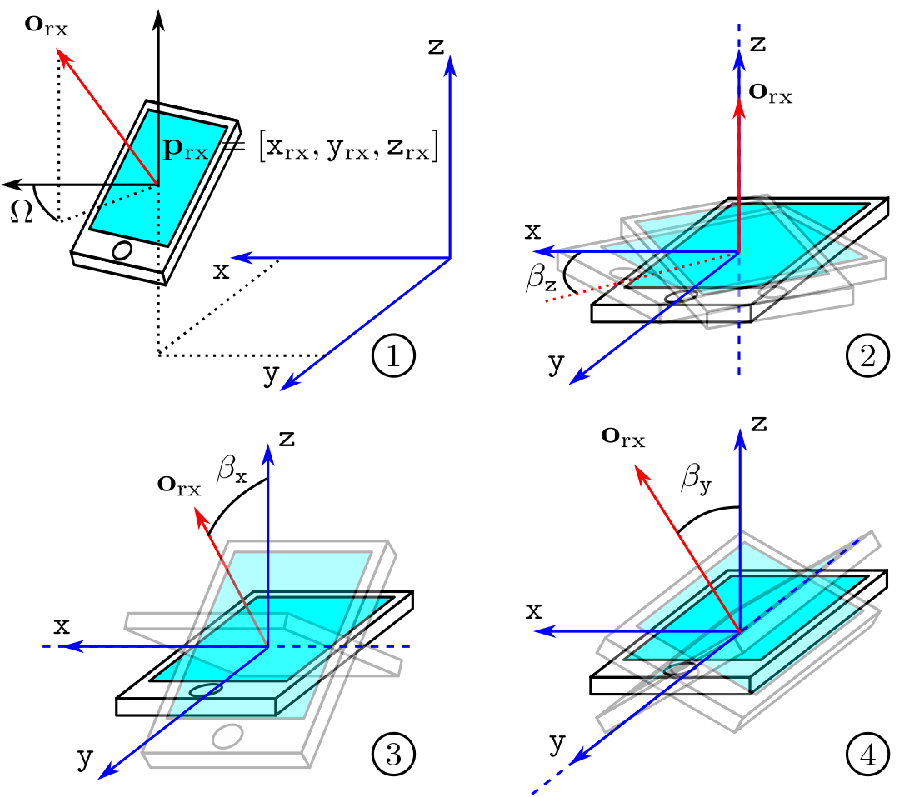}
		\caption{}
		\label{fig:rotation_ill}
	\end{subfigure}
	\caption{(a) Illustration of wavelength-dependent quantities. 1) incident angle $\psi=0\degree$ 2) incident angle $\psi=60\degree$. The central wavelength of the \ac{led} spectra and optical filter passbands at 0\degree incident angle are 437.5~nm, 512.5~nm, 587.5~nm and 662.5~nm. The optical filter passband width is 75~nm. (b) 1) UE position and horizontal orientation geometry. 2) UE rotation about $\mathtt{z}$-axis. 3) UE rotation about $\mathtt{x}$-axis. 4) UE rotation about $\mathtt{y}$-axis.}
	\label{fig:wavelength_domain_rotation_ill}
	\vspace{-7mm}
\end{figure}

Regarding the wavelength-domain characteristics, analytical spectrum models are used to improve the flexibility to configure the joint multiplexing system. The \ac{led} normalised spectral intensity can be defined by \cite{7728023}:
\begin{align}
\mathcal{S}_{n_{\rm t}}^{\rm led}(\lambda)=\frac{\frac{2}{\sqrt{\pi}}\exp\left(-\frac{\left(\lambda-\lambda_{n_{\rm t}}^{\rm led,c}\right)^2}{\Delta\lambda_{0.5}^2}\right)+\frac{4}{\sqrt{\pi}}\exp\left(-\frac{5(\lambda-\lambda_{n_{\rm t}}^{\rm led,c})^2}{\Delta\lambda_{0.5}^2}\right)}{\Delta\lambda_{0.5}\left(\frac{2+\sqrt{5}}{\sqrt{5}}+\mathrm{erf}\left(\frac{\lambda_{n_{\rm t}}^{\rm led,c}}{\Delta\lambda_{0.5}}\right)+\frac{2}{\sqrt{5}}\mathrm{erf}\left(\frac{\sqrt{5}\lambda_{n_{\rm t}}^{\rm led,c}}{\Delta\lambda_{0.5}}\right)\right)}, 
\label{eq:LED_spectrum}
\end{align}
where $\lambda_{n_{\rm t}}^{\rm led,c}$ is the central wavelength of the $n_{\rm t}$th \ac{led} and $\Delta\lambda_{0.5}$ is a parameter determining the spectrum shape of the \ac{led}, which is defined as:
\begin{align}
\Delta\lambda_{0.5}=\left\{\begin{array}{lr} \frac{5.5\mathcal{K}_{\rm B}T_{\rm j}}{\mathfrak{h}c}\left(\lambda_{n_{\rm t}}^{\rm led,c}\right)^2 &: \lambda_{n_{\rm t}}^{\rm led,c}\leq 560~\text{nm} \\
\frac{2.5\mathcal{K}_{\rm B}T_{\rm j}}{\mathfrak{h}c}\left(\lambda_{n_{\rm t}}^{\rm led,c}\right)^2 &:\lambda_{n_{\rm t}}^{\rm led,c}> 560~\text{nm} \end{array}
\right.,
\end{align}
where $\mathcal{K}_{\rm B}=1.38\times 10^{-23}$~J/K is the Boltzmann's constant, $T_{\rm j}=300$~K is the active layer temperature and $\mathfrak{h}=6.63\times 10^{-34}$~J/Hz is Planck's constant. Note that $\int_{\lambda_{\rm min}}^{\lambda_{\rm max}}\mathcal{S}^{\rm led}_{n_{\rm t}}(\lambda)\mathrm{d}\lambda=1$. This model has been demonstrated to be accurate compared to the off-the-shelf \ac{led} devices \cite{7728023}. The \ac{pd} spectral responsivity can be defined by the following expression \cite{hranilovic2006wireless}:
 \begin{align}
 \mathcal{R}^{\rm pd}(\lambda)=\frac{\eta_{\rm q}\mathfrak{q}\lambda}{\mathfrak{h}c},
 \label{eq:responsivity}
 \end{align}
where $\eta_{\rm q}$ is the quantum efficiency and $\mathfrak{q}=1.6\times 10^{-19}$~C is the electric charge. Considering a thin-film optical bandpass filter mounted on the $n_{\rm r}$th \ac{pd} with a central passband wavelength of $\lambda^{\rm of,c}_{n_{\rm r}}$ and a passband width of $\Delta\lambda^{\rm of}$ at $0$\degree incident angle, the spectral transmittance of the filter can be modelled by \cite{8546751}:
\begin{align}
\mathcal{G}_{n_{\rm r}}^{\rm of}(\lambda,\psi)=\left\{\begin{array}{lr} \mathcal{G}_{\rm T} &: \lambda_{n_{\rm r}}^{\rm of,l}(\psi)\leq\lambda\leq \lambda_{n_{\rm r}}^{\rm of,r}(\psi) \\
0 &: \text{otherwise} \end{array}
\right.,
\label{eq:optical_filter}
\end{align}
where $\mathcal{G}_{\rm T}$ is the transmittance of the optical filter, $\lambda_{n_{\rm r}}^{\rm of,l}(\psi)$ and $\lambda_{n_{\rm r}}^{\rm of,r}(\psi)$ are left and right edges of the filter passband, which are functions of the incident angle $\psi$:
\begin{align}
\lambda_{n_{\rm r}}^{\rm of,l}(\psi)=\left(\lambda^{\rm of,c}_{n_{\rm r}}-\Delta\lambda^{\rm of}/2\right)\sqrt{1-\sin^2\psi/\mathfrak{n}_{\rm e}^2}, \\
\lambda_{n_{\rm r}}^{\rm of,r}(\psi)=\left(\lambda^{\rm of,c}_{n_{\rm r}}+\Delta\lambda^{\rm of}/2\right)\sqrt{1-\sin^2\psi/\mathfrak{n}_{\rm e}^2},
\end{align}
and $\mathfrak{n}_{\rm e}$ is defined as the effective refraction index. 
Several examples of \eqref{eq:LED_spectrum}, \eqref{eq:responsivity} and \eqref{eq:optical_filter} have been depicted in Fig.~\ref{fig:wavelength_domain}. It is worth noting that the passbands of the four plotted optical filters shift to shorter wavelengths significantly when the light incident angle changes from $\psi=0\degree$ to $\psi=60\degree$. In indoor \ac{vlc}/\ac{lifi} applications, desired detector alignment is unlikely. Therefore, it is important to consider this bandpass shift characteristic. 


{\color{blue} \subsection{Receiver noise model}
Assuming that the background light power is negligible compared to the optical signal power, the dominant receiver noise components are the signal-dependent shot noise and the thermal noise. Therefore, the receiver noise variance can be calculated as:
\begin{align}
 {\sigma}_{{\rm rx},n_{\rm r}}^2=2\mathfrak{q}\tilde{i}_{n_{\rm r}}B_{\rm s}+\frac{4\mathcal{K}_{\rm B}\mathcal{T}_{\rm a}B_{\rm s}}{R_{\rm L}},
\end{align}
where $\tilde{i}_{n_{\rm r}}$ is the photocurrent of the $n_{\rm r}$th \ac{pd}, $B_{\rm s}=1/2T_{\rm s}$ is the signalling bandwidth and $\mathcal{T}_{\rm a}$ is the absolute temperature. For simplicity, the average photocurrent is used for the value of $\tilde{i}_{n_{\rm r}}$.
}

\section{\color{blue} Performance evaluation with various system configurations} 
\label{sec:system_configurations}

In Section~\ref{sec:spatial_wavelength_channel}, it has been shown that the \ac{mimo}-\ac{ofdm} channel is determined by many different parameters in the spatial and wavelength domains. In this section, we focus on exploring various system configurations. Specifically, we evaluate the performance of joint multiplexing systems with different sets of spatial and wavelength domain parameters, such as transmitter position $\mathbf{p}_{n_{\rm t}}^{\rm led}$ or optical filter passband $\Delta\lambda^{\rm of}$. {\color{blue} In a multiplexing system, the number of multiplexing channels has a significant impact on the aggregate data rate, which is determined by $\min(N_{\rm t},N_{\rm r})$. When evaluating the performance of joint multiplexing systems scaling with the number of \acp{led}/\acp{pd}, the data rate varies slightly if only one of the variables (number of \acp{led} or \acp{pd}) is changed. To avoid generating trivial results that are less meaningful, the numbers of \acp{led} and \acp{pd} are always identical ($N_{\rm t}=N_{\rm r}=I$) in the considered \ac{mimo}-\ac{ofdm} systems in the following sections.} For the convenience of description, we define a variable called the `number of elements' which is equivalent to the number of \acp{led}, \acp{pd} and data streams. In addition, to reduce the complexity of the metric evaluation, a uniform power allocation is used in this section: $q_{i}[k]=K/(K-2)$.

 

\subsection{\color{blue} Performance metric: average achievable rate with random \ac{ue} position and orientation}
Firstly, we define the average achievable rate as the performance metric. {\color{blue} Since the calculated \ac{snr} in \eqref{eq:SNR} corresponds to complex bipolar \ac{qam} symbols after a series of conversions, the achievable rate can be evaluated by \cite{9562203}}:
\begin{align}
C=\frac{1}{T_{\rm s}\left(K+N_{\rm cp}\right)}\sum_{i=1}^{I}\sum_{k=1}^{\tilde{K}}\log_2\left(1+\frac{\gamma_i[k]}{\Gamma}\right),
\label{eq:capacity1}
\end{align}
where $\gamma_i[k]$ is the \ac{snr} and $\Gamma$ is a gap factor between the achievable rate and the Shannon capacity, which is empirically configured to compensate the system impairment and imperfect constellation \cite{9562203}. Most of the parameters related to the \ac{snr} calculation are determined by the system configuration only, such as spectrum intensity or radiation pattern of an \ac{led}. However, the position and orientation of \acp{pd} are determined by the user behaviour. Therefore, the randomness in \ac{ue} position and orientation is considered in the evaluation of the achievable rate. 

The \ac{ue} geometry is depicted in Fig.~\ref{fig:rotation_ill}~1). The \ac{3d} position vector of the \ac{ue} is defined as $\mathbf{p}_{\rm rx}=[\mathtt{x}_{\rm rx},\mathtt{y}_{\rm rx},\mathtt{z}_{\rm rx}]$, where the $\mathtt{x}_{\rm rx}$ and $\mathtt{y}_{\rm rx}$ are the \ac{2d} coordinate of the \ac{ue} and $\mathtt{z}_{\rm rx}$ is the \ac{ue} height from the floor level, which is assumed to be deterministic in this study. It is also assumed that the size of the receiver is small, relative to the link distance, so that the multiple \acp{pd} mounted on the receiver are collocated. Thus, the position vector of the $n_{\rm r}$th \ac{pd} is the same as the \ac{ue} position: $\mathbf{p}_{n_{\rm r}}^{\rm pd}=\mathbf{p}_{\rm rx}$. In addition, an azimuth angle $\Omega$ is defined to determine the horizontal direction of the \ac{ue}. {\color{blue} Since there is no bias on the user position and the azimuth angle, it is assumed that these quantities are uniformly distributed, which have been widely accepted in the wireless communication research community:} $\mathtt{x}_{\rm rx}\sim\mathcal{U}_{(0,W_{\rm room})}$, $\mathtt{y}_{\rm rx}\sim\mathcal{U}_{(0,L_{\rm room})}$ and $\Omega\sim\mathcal{U}_{(0,2\pi]}$, where $\mathcal{U}_{\mathcal{I}}$ refers to the continuous uniform distribution with an interval $\mathcal{I}$. The notations $L_{\rm room}$ and $W_{\rm room}$ correspond to the length and width of the room. In addition to the random user position and horizontal direction, two user device orientation scenarios are considered in this study. The random orientation of the \ac{ue} is defined by three rotation angles: $\beta_{\mathtt{z}}$, $\beta_{\mathtt{x}}$ and $\beta_y$ which correspond to the yaw, pitch and roll rotations about the $\mathtt{z}$, $\mathtt{x}$ and $\mathtt{y}$-axes, as illustrated in Fig.~\ref{fig:rotation_ill}~2), 3) and 4). Then, the \ac{pd} orientation vector $\mathbf{o}_{n_{\rm r}}^{\rm pd}$ is determined by:
\begin{align}
{\left[\mathbf{o}_{n_{\rm r}}^{\rm pd}\right]}^{\mathrm{T}}=\mathbf{R}_{\mathtt{z}}\mathbf{R}_{\mathtt{x}}\mathbf{R}_{\mathtt{y}}\left[\mathbf{o}_{n_{\rm r}}^{\rm pd,\uparrow}\right]^{\mathrm{T}},
\label{eq:orientation_rotation}
\end{align}
where $\mathbf{o}_{n_{\rm r}}^{\rm pd,\uparrow}$ is the \ac{pd} orientation vector of the $n_{\rm r}$th \ac{pd} when the \ac{ue} has an upward facing orientation with the normal vector of $\mathbf{o}_{\rm rx}=[0,0,1]$; $\mathbf{R}_{\mathtt{z}}$, $\mathbf{R}_{\mathtt{x}}$ and $\mathbf{R}_{\mathtt{y}}$ are \ac{3d} rotation matrices corresponding to $\beta_{\mathtt{z}}$, $\beta_{\mathtt{x}}$ and $\beta_y$, respectively. They are defined by:
\begin{table}[!t]
	\caption{Upward and random orientations rotation angle characteristics}
	\centering
	\footnotesize
	\begin{tabular}{c|c||c c}
		\hline
		~ &(a) upward & \multicolumn{2}{c}{(b) random} \\		
		\hline
		~ & value & mean $\mu$ & standard deviation $\sqrt{2}b$ \\
		\hline
		$\beta_{\mathtt{z}}$ & $\Omega-90\degree$  & $\Omega-90\degree$ & $3.67\degree$ \\
		$\beta_{\mathtt{x}}$ & $0\degree$  & $40.78\degree$ & $2.39\degree$ \\
		$\beta_{\mathtt{y}}$ & $0\degree$  & $-0.84\degree$ & $2.21\degree$ \\
		\hline
	\end{tabular}
	\label{table:rx_orientation}
\end{table}
\begin{align}
&\mathbf{R}_{\mathtt{z}}=\left[
\begin{matrix}
\cos\beta_{\mathtt{z}} & -\sin\beta_{\mathtt{z}} & 0 \\ \sin\beta_{\mathtt{z}} & \cos\beta_{\mathtt{z}} & 0 \\
0 & 0 & 1
\end{matrix}\right],~\mathbf{R}_{\mathtt{x}}=\left[
\begin{matrix}
1 & 0 & 0 \\ 0 & \cos\beta_{\mathtt{x}} & -\sin\beta_{\mathtt{x}} \\
0 & \sin\beta_{\mathtt{x}} & \cos\beta_{\mathtt{x}}
\end{matrix}\right],~\mathbf{R}_{\mathtt{y}}=\left[
\begin{matrix}
\cos\beta_{\mathtt{y}} & 0 & \sin\beta_{\mathtt{y}} \\ 0 & 1 & 0 \\
-\sin\beta_{\mathtt{y}} & 0 & \cos\beta_{\mathtt{y}}
\end{matrix}\right].
\end{align} 
In the first scenario, an upward facing \ac{ue} is considered, which corresponds to large devices such as laptops. In this case, we have $\beta_{\mathtt{z}}=\Omega-90\degree$ and $\beta_{\mathtt{x}}=\beta_{\mathtt{y}}=0\degree$, as concluded in Table~\ref{table:rx_orientation}~(a). In the second scenario, hand-held devices are considered with a random orientation. The statistics of $\beta_{\mathtt{z}}$, $\beta_{\mathtt{x}}$ and $\beta_{\mathtt{y}}$ have been obtained via experimental measurement and modelled using Laplace and Gaussian distributions \cite{8790655}. For sitting users, the three rotation angles follow Laplace distributions: $\beta\sim\mathcal{L}(\mu,b)$ with a mean value of $\mu$ and a standard deviation of $\sqrt{2}b$. The means and standard deviations of different rotation angles are concluded in Table~\ref{table:rx_orientation}~(b). By incorporating the \ac{ue} randomness, the average achievable rate can be evaluated by:
\begin{align}
 \bar{C}=\mathbb{E}_{\mathtt{x}_{\rm rx},\mathtt{y}_{\rm rx},\Omega}\left[C\right],~\bar{C}=\mathbb{E}_{\mathtt{x}_{\rm rx},\mathtt{y}_{\rm rx},\Omega,\beta_{\mathtt{z}},\beta_{\mathtt{x}},\beta_{\mathtt{y}}}\left[C\right],
 \label{eq:average_capacity}
\end{align}
for the cases of upward and random orientation scenarios, respectively. Due to the complexity issues, these metrics have to be evaluated using a Monte Carlo approach.

\subsection{\color{blue} System configuration with BADS algorithm} 
\label{subsec:configuration_methodology}

{\color{blue} In this paper, the average achievable rates \eqref{eq:average_capacity} with several sets of empirically selected parameters are evaluated. In addition, we aim to evaluate the maximum potential of the proposed joint multiplexing system in terms of achievable rate. With the number of parameters exceeding ten, it is very time-consuming to find the best system configuration parameters with an exhaustive search approach by testing all combinations of different parameter values. Therefore, a suitable optimisation tool is used to efficiently search the desired system configuration.

The optimisation objective function \eqref{eq:average_capacity} is a complex function of the considered system parameters by equations from \eqref{eq:SNR} to \eqref{eq:capacity1}. Additionally, the numerical evaluation of \eqref{eq:average_capacity} requires averaging over random user orientations and positions with a Monte-Carlo approach. This leads to a noisy, high dimensional and high complexity objective function, which cannot be solved by a conventional gradient-based optimisation method, such as linear programming. Instead, a state-of-the-art black-box optimisation tool is used to tackle this problem. Black-box optimisation methods are used when the objective function is unknown, non-smooth or complicated to evaluate, where the derivate of the objective function is unavailable, unreliable or impractical to calculate \cite{audet2017derivative}. The used optimisation is known as \ac{bads} which combines the capability of Bayesian optimisation in optimising expensive and noisy black-box functions with the low computational cost of \ac{mads}. It has been demonstrated that the \ac{bads} algorithm outperforms a number of widely used and state-of-the-art non-convex, derivative-free optimisation algorithms on many practical problems \cite{acerbi2017practical}. This type of optimisation methods have wide applications in the fields of computational neuroscience and machine learning where models are evaluated via stochastic simulation or numerical approximation \cite{li2020quantum}.

}
\begin{algorithm}[t!]
	\fontsize{10pt}{10pt}\selectfont
	\SetAlgoLined
	\textbf{\textit{Initialise:}} $n_{\rm iter}=0$, $\Delta^{\rm m}_{0}=2^{-10}$, $\Delta^{\rm p}_{0}=1$, evaluate $\mathfrak{f}(\mathbf{x}_0)$ \\
	\While{$n_{\rm iter}\leq N_{\rm iter,max}$ and $\Delta^{\rm p}_{0}>\Delta^{\rm p}_{\rm th}$}{
		\Repeat{$\mathfrak{f}(\mathbf{x}_{\rm search})$ provides suffient improvement or reaches maximum search number $N_{\rm search}$}{generate search input $\mathbf{x}_{\rm search}$ and evaluate $\mathfrak{f}(\mathbf{x}_{\rm search})$}
		\If{Search stage is unsuccessful}{
			generate poll set $P_{n_{\rm iter}}$, sort it using the acquisition function and
			evaluate $\mathfrak{f}(\mathbf{x})$ on $\mathbf{x}\in P_{n_{\rm iter}}$}
		\eIf{$n_{\rm iter}$th iteration is successful}{
			update incumbent $\mathbf{x}_{n_{\rm iter}+1}$ \\
			\If{Poll stage is successful}{$\Delta^{\rm m}_{n_{\rm iter}+1}=2\Delta^{\rm m}_{n_{\rm iter}}$, $\Delta^{\rm p}_{n_{\rm iter}+1}=2\Delta^{\rm p}_{n_{\rm iter}}$}
		}{$\Delta^{\rm m}_{n_{\rm iter}+1}=\frac{1}{2}\Delta^{\rm m}_{n_{\rm iter}}$, $\Delta^{\rm p}_{n_{\rm iter}+1}=\frac{1}{2}\Delta^{\rm p}_{n_{\rm iter}}$}
		$n_{\rm iter}=n_{\rm iter}+1$
	}
	\caption{Bayesian Adaptive Direct Search \cite{acerbi2017practical}}
	\label{algo:BADS}
\end{algorithm} 
The major operations in the \ac{bads} algorithm are introduced in Algorithm~\ref{algo:BADS}. The \ac{bads} algorithm considers a possibly noisy objective function $\mathfrak{f}(\mathbf{x})$ with an $N_{\rm d}$ dimension input variable vector $\mathbf{x}\in\mathbb{R}^{N_{\rm d}}$. A lower bound vector $\mathbf{x}_{\rm LB}$ and an upper bound vector $\mathbf{x}_{\rm UB}$ must be provided as the constraints to the variable vector $\mathbf{x}$. An optional plausible bound vectors $\mathbf{x}_{\rm PLB}$ and $\mathbf{x}_{\rm PUB}$ can also be provided to indicate the regions where the optimal solutions are more likely. At the beginning of the algorithm, an initial solution $\mathbf{x}_{0}$ will be provided as a starting point. Then, the algorithm enters the search stage and executes a series of efficient local Bayesian optimisations around the point and try to find a solution which is better than the current one. Each new evaluation of the cost function is used as a new sample of the data set to re-train a Gaussian process model to be a more accurate surrogate. If the algorithm fails to find a better solution in the search stage, it means the current Gaussian process model is not useful in the optimisation. Then, the algorithm can switch to the poll stage which uses model-free opportunistic optimisation to compensate. The notations $\Delta^{\rm m}_{n_{\rm iter}}$ and $\Delta^{\rm p}_{n_{\rm iter}}$ are the mesh and poll size of the $n_{\rm iter}$th iteration. They function as the adaption step sizes, which vary in the optimisation process. In unsuccessful iterations, $\Delta^{\rm m}_{n_{\rm iter}}$ and $\Delta^{\rm p}_{n_{\rm iter}}$ are decreased to increase the searching granularity. In successful iteration in the poll stage, they are increased to speed up the convergence speed. Finally, the algorithm ends when the poll size is smaller than a predefined threshold or the maximum number of iterations is reached. 

{\color{blue} More details about the implementation, convergence analysis, optimality condition and complexity of \ac{bads} can be found in \cite{acerbi2017practical,audet2006mesh}. In this study, we have used the \ac{bads} algorithm implementation in MATLAB provided by the authors of \cite{acerbi2017practical}. It is worth noting that the \ac{bads} algorithm is a semi-local algorithm which may generate a global or a local optimal solution. To increase the chance of obtaining global optimal solutions, a multi-start strategy is used and the number of iterations with different initial solutions for each optimisation problem equals or greater than ten. In this study, the running time of executing the \ac{bads} algorithm with multiple iterations varies from a few seconds to tens of minutes with the increase of element number $I$. However, the obtained system configuration parameters from searching the problems with the \ac{bads} algorithm are static and used to implement the \ac{vlc} joint multiplexing system. These obtained parameters are fixed once the system is implemented and will not change during the operation of the system.	Therefore, unlike many signal processing algorithms, there is little computational complexity concerns about the use of \ac{bads} algorithm in this work.
}


\begin{figure}[t!]
	\centering
	\begin{subfigure}{.45\textwidth}
		\centering
		\includegraphics[width=.95\linewidth]{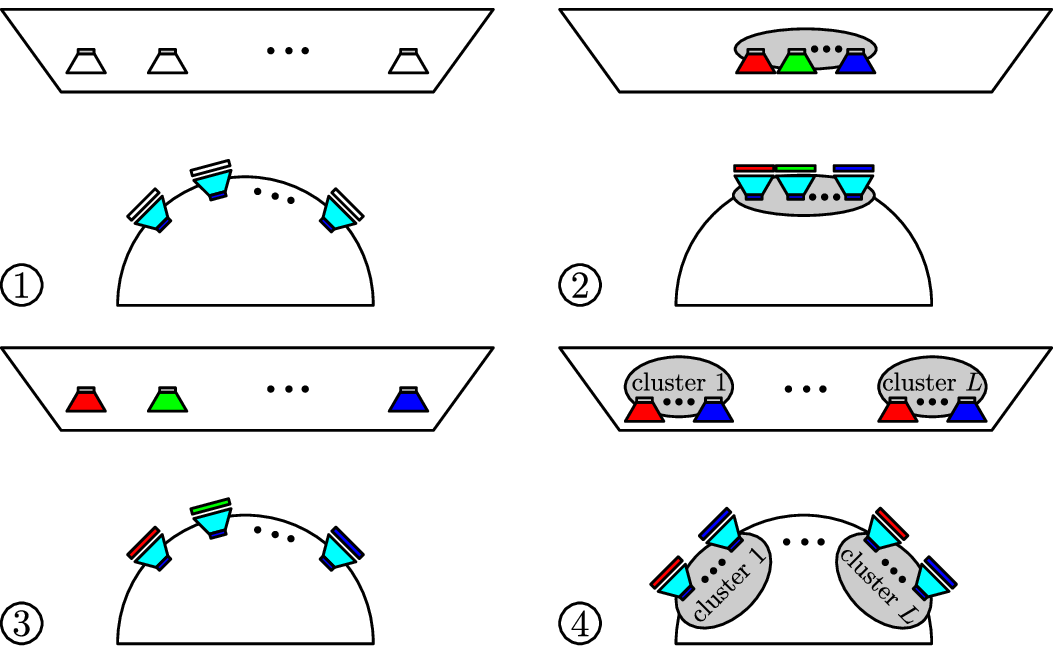}
		\caption{}
		\label{fig:configuration_scenarios}
	\end{subfigure}%
	\begin{subfigure}{.55\textwidth}
		\centering
		\includegraphics[width=.95\linewidth]{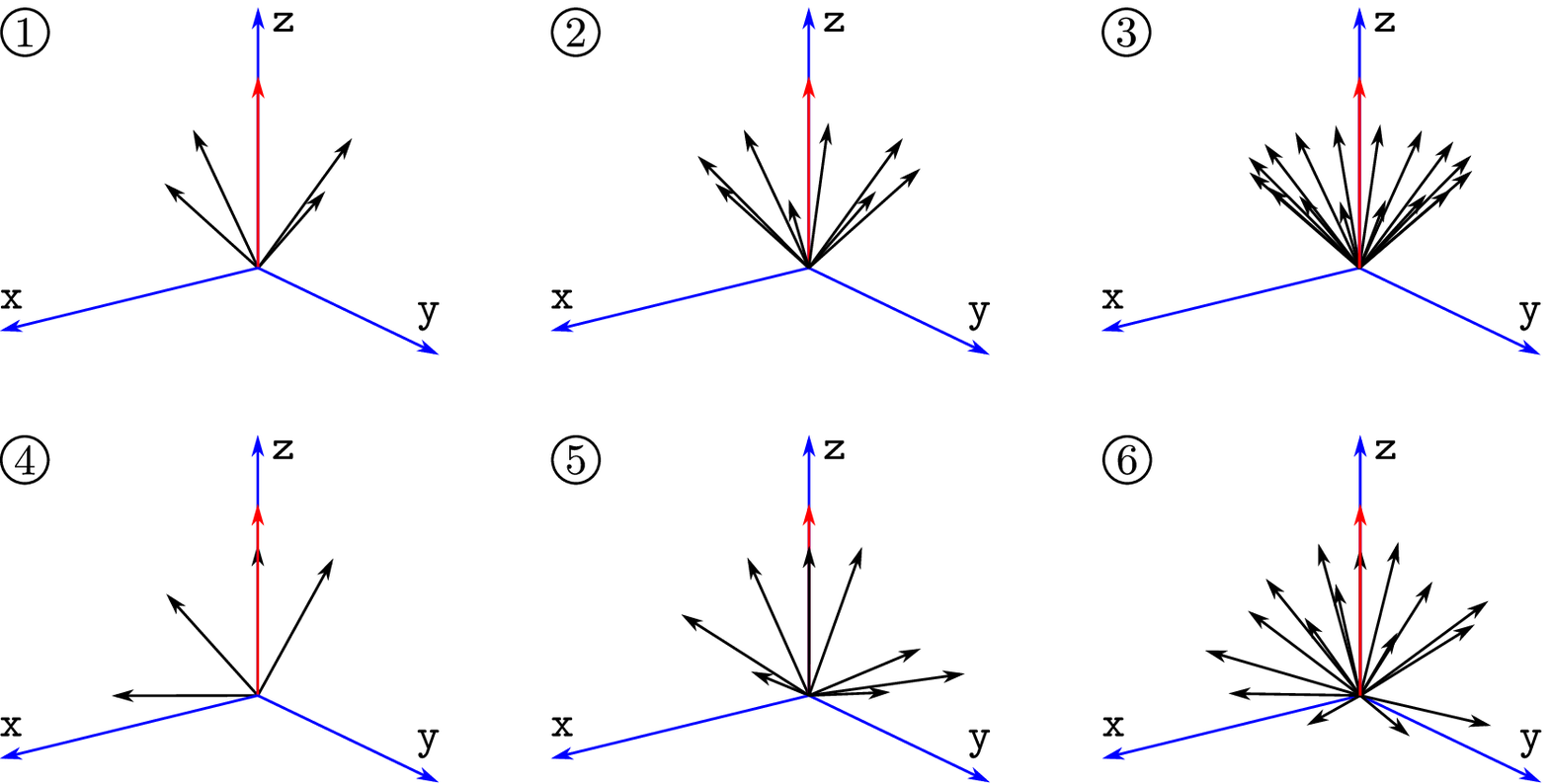}
		\caption{}
		\label{fig:angular_diversity_Rx}
		\vspace{-7mm}
	\end{subfigure}
	\caption{(a) Illustration of various LED layouts on the ceiling and PD orientations on the mobile side with different configuration strategies: 1) Division in space domain. 2) Division in wavelength domain. 3) Division in both spatial and wavelength domains. 4) Division in either spatial or wavelength domain. (b) Angular diversity receivers: 1), 2) and 3) show three PR examples with $N_{\rm r}=4,8,16$. 4), 5) and 6) show three HR examples with $N_{\rm r}=4,8,16$.}
	\label{}
\end{figure}

\subsection{\color{blue} Overview of system configuration strategies}
Now we consider four potential configuration strategies: 1) division in the space domain, 2) division in the wavelength domain, 3) division in both the spatial and wavelength domains, 4) division in either the spatial or wavelength domain, as demonstrated in Fig.~\ref{fig:configuration_scenarios}. The first two strategies use the \acp{dof} in the spatial or wavelength domain only, which are equivalent to \ac{smx} and \ac{wdm} transmission systems, respectively. These two strategies are referred as \ac{sd} and \ac{wd}. They will be covered in Section~\ref{subsec:SMP} and~\ref{subsec:WDM}, respectively. The third and fourth strategies use the \acp{dof} in the spatial and/or wavelength domain jointly. The third strategy considers each optical element has a distinct spatial and wavelength domain feature, as illustrated in Fig.~\ref{fig:configuration_scenarios}~3). This strategy leads to excessive decorrelation of the channel matrix at the cost of inefficient use of spatial and wavelength \acp{dof}. In contrast, the fourth strategy considers to let each optical element to have a unique feature in either the spatial or wavelength domain, as illustrated in Fig.~\ref{fig:configuration_scenarios}~4). It is expected to provide enough channel matrix decorrelation and also be able to support a greater number of multiplexing channels compared to the \ac{sd} and \ac{wd} strategies. This joint strategy will be covered in Section~\ref{subsec:HMP}.

\subsection{Spatial division strategy}
\label{subsec:SMP}
Firstly, the configurable parameters considered in \ac{sd} strategy is introduced. Since only the \ac{dof} in spatial domain is utilised for multiplexing transmission, wavelength domain parameters are empirically selected with identical values for all \acp{led} and optical filters: $\Delta\lambda^{\rm of}=300$~nm, $\lambda_{n_{\rm t}}^{\rm led,c}=550$~nm for $n_{\rm t}=1,2,\cdots,N_{\rm t}$ and $\lambda_{n_{\rm r}}^{\rm of,c}=550$~nm for $n_{\rm r}=1,2,\cdots,N_{\rm r}$. On the transmitter side, the position vector of the $n_{\rm t}$th \ac{led} is defined as:
\begin{align}
\mathbf{p}_{n_{\rm t}}^{\rm led}=[\mathtt{x}_{n_{\rm t}},\mathtt{y}_{n_{\rm t}},\mathtt{z}_{\rm tx}],
\end{align}
where $\mathtt{z}_{\rm tx}$ is the height coordinate of the \acp{led} determined by the height of the room; $\mathtt{x}_{n_{\rm t}}$ and $\mathtt{y}_{n_{\rm t}}$ are the \ac{2d} horizontal coordinates of the \ac{led}, which are configurable parameters. Regarding the orientation of \acp{led}, a straight downward orientation of $\mathbf{o}_{n_{\rm t}}^{\rm led}=[0,0,-1]$ is used to guarantee the lighting performance and a wide coverage. Finally, the radiation pattern is defined by the \ac{led} half-power semiangle $\phi_{1/2}$, which is another important configurable parameter. On the receiver side, it is impractical to control the exact position and orientation of each \ac{pd} due to the random user location and orientation, while the \ac{fov} coefficient $m_{\rm fov}$ can be manipulated to control the light reception pattern, which is a configurable parameter of interest. On the other hand, the small \ac{ue} size assumption prevents the effective spatial decorrelation using different \ac{pd} positions. In order to achieve low spatial correlation on the receiver side, two types of angular diversity receivers are considered in this study: \ac{pr} and \ac{hr} \cite{7109107}. In equation \eqref{eq:orientation_rotation}, the upward facing \ac{pd} orientation vector can be represented by $\mathbf{o}_{n_{\rm r}}^{\rm pd,\uparrow}=\left[\cos\omega_{n_{\rm r}}\sin\theta_{n_{\rm r}},\sin\omega_{n_{\rm r}}\sin\theta_{n_{\rm r}},\cos\theta_{n_{\rm r}}\right]$,
where $\omega_{n_{\rm r}}$ is a \ac{pd} azimuth angle and  $\theta_{n_{\rm r}}$ is a \ac{pd} elevation angle. In a \ac{pr}, the values of $\theta_{n_{\rm r}}$ is defined by a \ac{pr} elevation angle $\theta_{\rm pd}$, which is a suitable configurable parameter. The \ac{pd} azimuth angles are specified as:
\begin{align}
\omega_{n_{\rm r}}=\frac{2\pi}{N_{\rm r}}(n_{\rm r}-1).
\label{eq:pr_angle}
\end{align}
In a \ac{hr}, the elevation angles are specified as:
\begin{align}
\theta_{n_{\rm r}}=\arccos\left(s_{n_{\rm r}}\right),
\label{eq:hr_angle1}
\end{align}
with  $s_{n_{\rm r}}=1-\frac{2(n_{\rm r}-1)}{2N_{\rm r}-1}$, and the azimuth angles are specified as:
\begin{align}
\omega_{n_{\rm r}}=\left(\omega_{n_{\rm r}-1}+3.6\left(2N_{\rm r}\left(1-s_{n_{\rm r}}^2\right)\right)^{-1/2}\right)\mathrm{mod}~2\pi,
\label{eq:hr_angle2}
\end{align}
for $n_{\rm r}=2,\cdots,N_{\rm r}$ and $\omega_{1}=0$. Fig.~\ref{fig:angular_diversity_Rx} demonstrate six examples of \acp{pr} and \acp{hr} with $N_{\rm r}=4,8,16$. In summary, the considered configurable parameters in the \ac{sd} strategy include: $\mathtt{x}_{n_{\rm t}}$, $\mathtt{y}_{n_{\rm t}}$, $\phi_{1/2}$, $m_{\rm fov}$, $(\theta_{\rm pd})$. Note that $\theta_{\rm pd}$ is only considered when \acp{pr} are used.

\begin{figure}[!t]
	\begin{center}
		\includegraphics[width=1\textwidth]{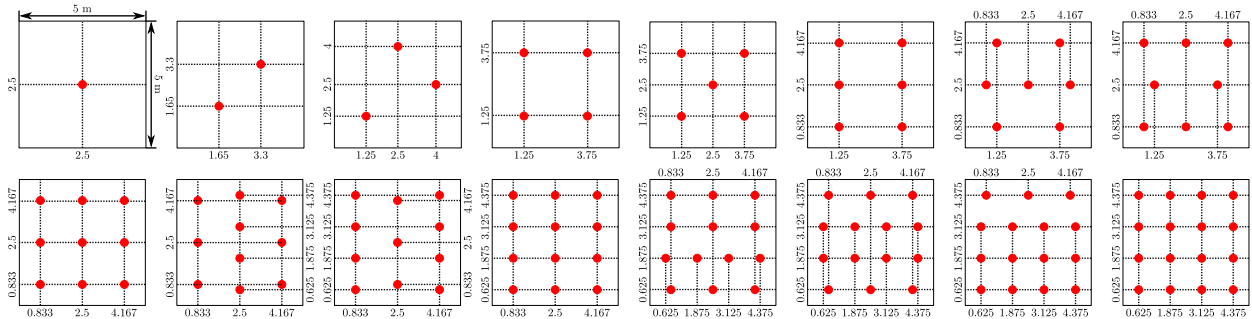}
	\end{center}
	\caption{Empirical configuration to \ac{2d} \ac{led} position layout.}
	\label{fig:LED_layout}
	\vspace{-7mm}
\end{figure}

\subsubsection{Empirical configuration}
In cases with the empirical configuration, the values of $\mathtt{x}_{n_{\rm t}}$ and $\mathtt{y}_{n_{\rm t}}$ with different number of $N_{\rm t}$ are specified according to the \ac{2d} layouts shown in Fig.~\ref{fig:LED_layout}. The intuition of this \ac{led} spatial distribution is to distribute the \acp{led} with wider separations to guarantee a low spatial correlation on the transmitter side. On the other hand, the \acp{led} are distributed in a manner to let the system cover the entire room. An \ac{led} halfpower semiangle of $\phi_{1/2}=60\degree$ is used to ensure a wide coverage of each \ac{led}. An \ac{fov} coefficient of $m_{\rm fov}=1.4738$ is used as this value has been found to achieve a good match to practical \ac{pd} characteristics \cite{7109107}. In the case of \ac{pr}, a \ac{pd} elevation angle of $\theta_{\rm pd}=40\degree$ is used as it has been demonstrated to achieve a high data rate in a \ac{mimo} system \cite{7109107}.


\subsubsection{Configuration with \ac{bads} algorithm}
\label{subsubsec:SD_optimised}
{\color{blue} It is straightforward to use the \ac{bads} algorithm introduced in Section~\ref{subsec:configuration_methodology} to search system configuration parameters of interest with the following procedures: 1. Formulate a general black-box optimisation problem with the considered configurable parameters as input variables $\mathbf{x}$ and average achievable data rate $\bar{C}$ as the objective function. 2. Define bound vectors based on the practical constraints of the considered input variables. 3. Generate an initial solution vector $\mathbf{x}_0$, where each variable is randomly selected from the values between the plausible bounds with equal probability. 4. Execute the \ac{bads} algorithm multiple times with different initial solutions. The black-box optimisation problem for system configurations with the \ac{sd} strategy is defined as: 
}
\begin{align}
& \underset{\mathcal{X}}{\text{maximise}} &&\bar{C}^{\rm SD}(\mathcal{X})~\text{with}~\mathcal{X}=\left\{\mathtt{x}_{n_{\rm t}},\mathtt{y}_{n_{\rm t}},\phi_{1/2},m_{\rm fov},(\theta_{\rm pd})\right\} \\
&\text{subject to} &&0<\frac{\mathtt{x}_{n_{\rm t}}}{W_{\rm room}}<1~\text{for}~n_{\rm t}=1,\cdots,I, \\
& ~ && 0<\frac{\mathtt{y}_{n_{\rm t}}}{L_{\rm room}}<1~\text{for}~n_{\rm t}=1,\cdots,I, \\
& ~ && 0\degree<\phi_{1/2}\leq 60\degree, \\
& ~ && m_{\rm fov}\geq 1, \\
& ~ && (0\degree\leq\theta_{\rm pd}\leq 90\degree).
\end{align}
Note that $\theta_{\rm pd}$ only exist in the optimisation with \ac{pr}.
\begin{table}[!t]
	\caption{}
	\centering
	\begin{tabular}{c c c|c c c}
		\hline
		\hline
		Parameters & Symbol & Values & Parameters & Symbol & Values \\
		\hline
		Room width/length/height~[m] & $W_{\rm room}/L_{\rm room}/H_{\rm room}$ & $\rm 5/5/3$ & DCO-OFDM clipping level & $\kappa$ & 3.2 \\
		Total optical power~[W] & $\sum_{n_{\rm t}=1}^{N_{\rm t}}p_{n_{\rm t}}$ & 80 & PD physical area~[$\rm cm^2$] & $A_{\rm pd}$ & $1$ \\
		Height of LED/PD~[m] & $\mathtt{z}_{\rm tx}/\mathtt{z}_{\rm rx}$ & 3/0.75 & Number of subcarriers & $K$ & 256 \\
		Wavelength range~[nm] & $\lambda_{\rm min}/\lambda_{\rm max}$ & 400/700 & Modulation bandwidth~[MHz] & $1/2T_{\rm s}$ & 50 \\
		Optical filter transmittance & $\mathcal{G}_{\rm T}$ & 1 & LED/PD 3dB bandwidth~[MHz] & $f_{\rm led,c}$ & $35/106$ \\
		PD quantum efficiency & $\eta_{\rm q}$ & 0.8 & Modulation gap factor & $\Gamma$ & 6.06~dB \\
		Effective index & $\mathfrak{n}_{\rm e}$ & 2 & Cyclic-prefix & $N_{\rm cp}$ & 30 \\
		\hline
		\hline
	\end{tabular}
	\label{table:Parameters}
\end{table}

\begin{figure}[t!]
	\centering
	\begin{subfigure}{.5\textwidth}
		\centering
		\includegraphics[width=1\linewidth]{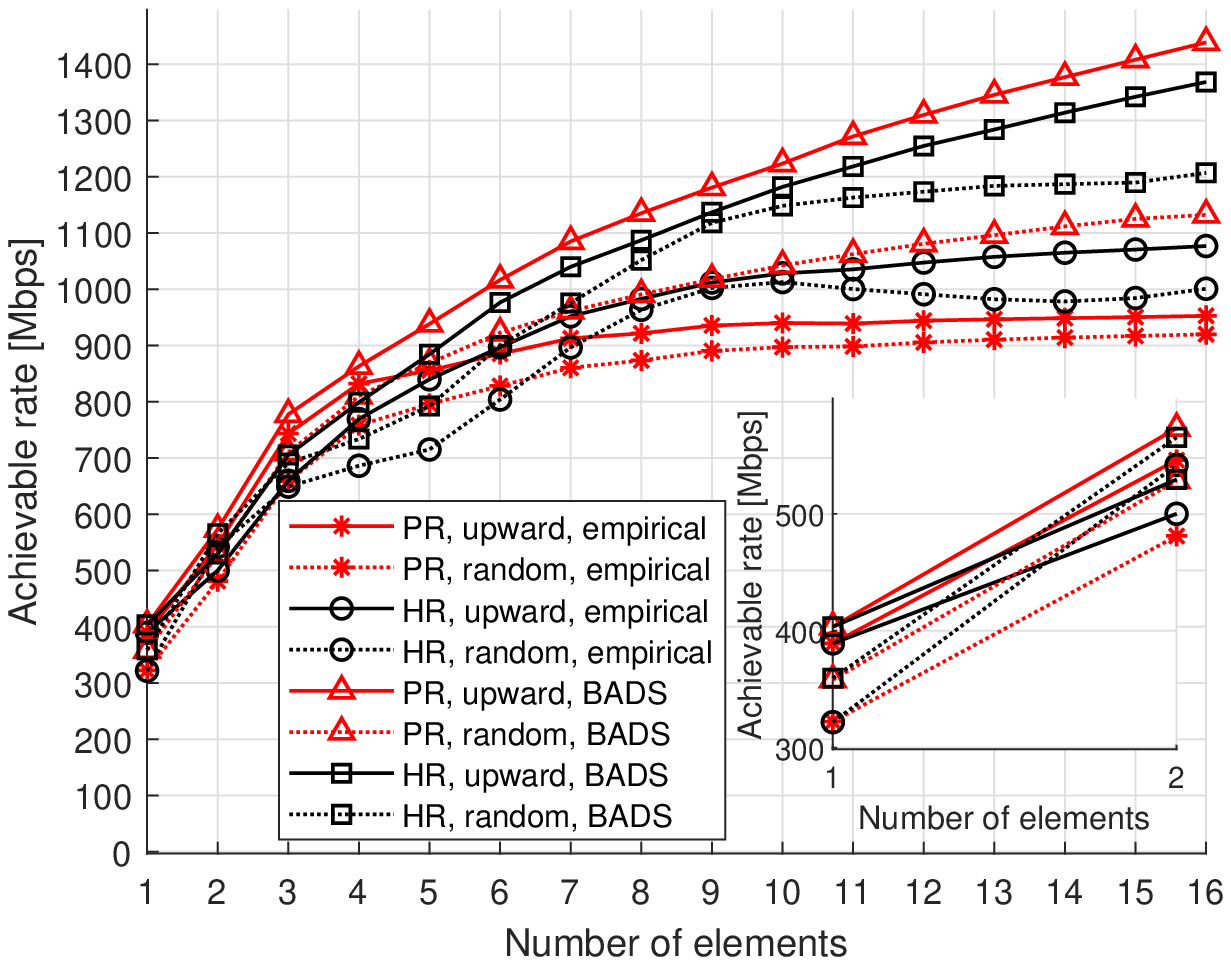}
		\caption{}
		\label{fig:SE_SMP}
	\end{subfigure}%
	\begin{subfigure}{.5\textwidth}
		\centering
		\includegraphics[width=1\linewidth]{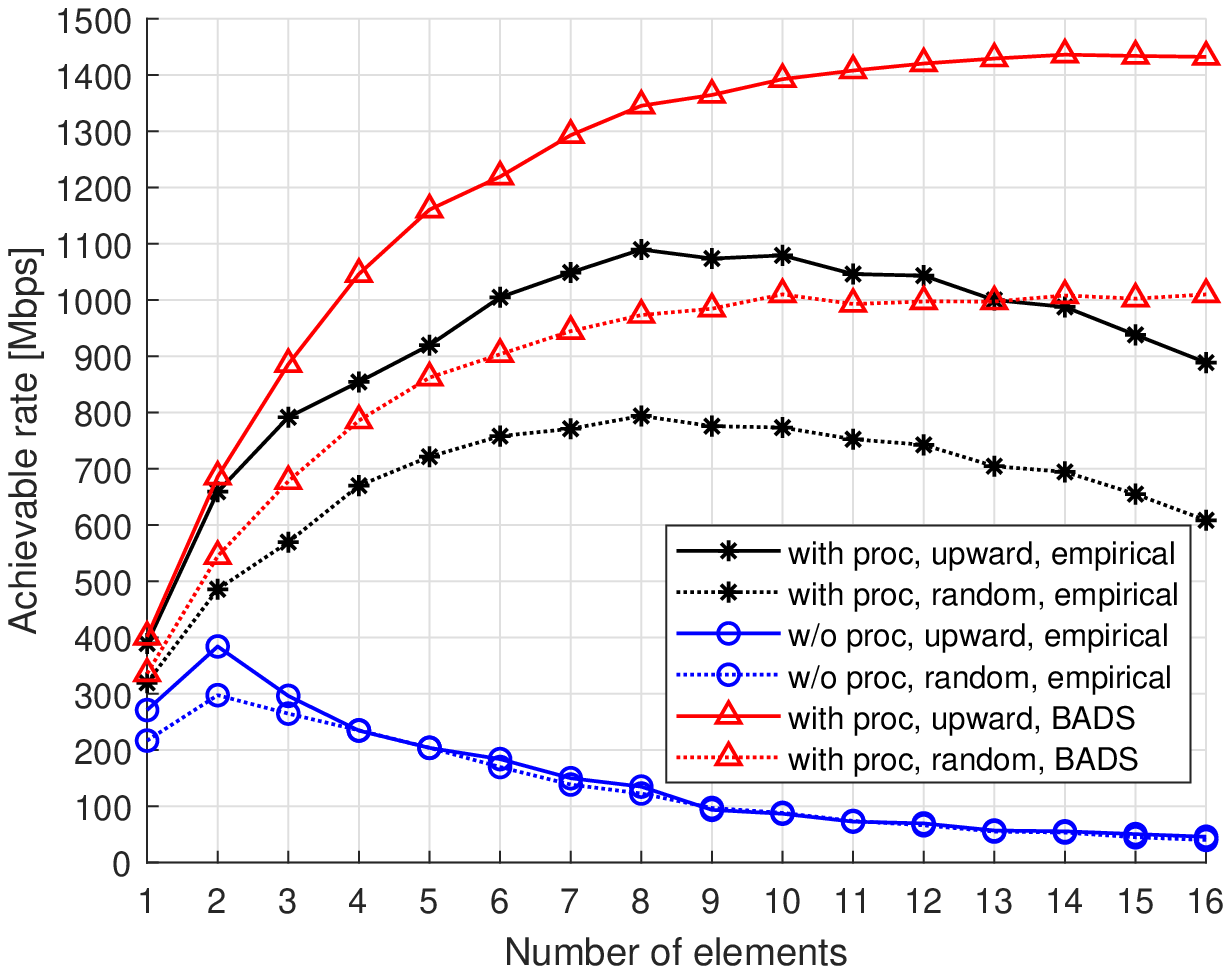}
		\caption{}
		\label{fig:SE_WDM}
	\end{subfigure}
	\caption{Average achievable rate against number of elements achieved by joint multiplexing systems with (a) \ac{sd} strategy and (b) \ac{wd} strategy. 'upward' and 'random' refer to the upward facing receiver scenario and random orientation scenario, respectively. `with proc' and `w/o proc' refer to with \ac{mimo} processing and without \ac{mimo} processing, respectively.}
	\label{}
	\vspace{-7mm}
\end{figure}

Fig.~\ref{fig:SE_SMP} shows the average achievable rate results against the number of elements (up to 16) achieved by systems with \ac{sd} strategy. The remaining system parameters (irrelevant to \ac{dof} in the spatial domain) are listed in Table~\ref{table:Parameters}. {\color{blue} Regarding the gap factor value, it has been shown in \cite{8901160} that a gap factor of 6~dB is sufficient for $M$-\ac{qam} modulation to achieve a \ac{ber} at $1\times 10^{-3}$ for $M=2$. With an increase in $M$, the required gap factor decreases. This implies that by using a fixed gap factor of 6~dB is sufficient for the calculated achievable rate with a \ac{ber} of $1\times 10^{-3}$.} Note that as the number of elements changes, the total transmission optical power stays the same and is equally distributed to each \ac{led}. Both cases with a \ac{pr} and \ac{hr} are presented. In addition, the cases with upward facing/random orientation scenarios and with empirical/optimised configurations are demonstrated. With an increase in the number of elements, the average achievable rate also increases due to more available multiplexing channels. However, a further increase in the number of elements (beyond 5 to 7) leads to a less average achievable rate improvement, especially for the cases with empirical configurations using \ac{pr}. This is because with a greater number of elements, the channel spatial correlation also increases. This leads to a channel matrix with worse channel conditions, which achieves fewer and weaker multiplexing channels. Intuitively, the performance of systems with optimised configurations are better than those with empirical configurations. With 16 elements, the average achievable rates by empirically configured systems are lower than 1100~Mbps, while the systems with optimised configurations achieved a sum rate in the range of 1100 to 1500~Mbps. On the other hand, systems with random orientation scenarios received a performance penalty compared to cases with upward facing receivers due to the more severe misalignment. Regarding different types of angular diversity receivers, the \acp{pr} offer better performance with a fewer number of elements, while the \ac{hr} offers a slightly better performance with a large numbers of elements.


\subsection{Wavelength division strategy}
\label{subsec:WDM}

In cases employing the \ac{wd} strategy, the joint multiplexing system is equivalent to a conventional \ac{wdm} system with \ac{mimo} processing techniques, which is similar to those in \cite{8617505}. All spatial domain parameters are empirically defined as follows: all \acp{led} are located in the centre of the room $\mathbf{p}_{n_{\rm t}}^{\rm led}=[W_{\rm room}/2,L_{\rm room}/2,\mathtt{z}_{\rm tx}]$ with an orientation of $\mathbf{o}_{n_{\rm t}}^{\rm led}=[0,0,-1]$ and all \acp{pd} have the same orientation of $\mathbf{o}_{n_{\rm r}}^{\rm pd,\uparrow}=[0,0,1]$. The wavelength domain configurable parameters of interest in cases with the \ac{wd} strategy include the \ac{led} central wavelength $\lambda_{n_{\rm t}}^{\rm led,c}$, optical filter passband centre $\lambda_{n_{\rm r}}^{\rm of,c}$ and optical filter passband width $\Delta\lambda^{\rm of}$.

\subsubsection{With and without \ac{mimo} processing}
With the \ac{wd} strategy, the precoding and post-detection process transform the `colour'-\ac{mimo} channel matrix with inter-colour crosstalk to a diagonal matrix. This operation is expected to mitigate the excessive crosstalk caused by the bandpass shift phenomenon. To highlight this important feature, the results of conventional \ac{wdm} systems without \ac{mimo} processing are also included for comparison. In the case of conventional \ac{wdm} systems, the precoding and post-detection matrices are set to identity matrices: $\mathbf{F}_k=\mathbf{W}_k=\mathbf{I}$, where $\mathbf{I}$ is defined as an identity matrix.

\subsubsection{Empirical configuration}
In the case of empirical configurations, the passband width of optical filters are defined by:
\begin{align}
\Delta\lambda^{\rm of}&=(\lambda_{\rm max}-\lambda_{\rm min})/I.
\label{eq:passband_width1}
\end{align}
The intuition of this setting is that the passband should be wide enough to accept most of the signal power from one \ac{led}, but not too wide to receive too much power from \acp{led} of other colours. In addition, the spectrum centre of the $i$th \ac{led} and the passband centre of the $i$th optical filter are defined by:
\begin{align}
 \lambda_{i}^{\rm led,c}&=\lambda_{i}^{\rm of,c}=\lambda_{\rm min}+(i-0.5)\Delta\lambda^{\rm of}.
 \label{eq:spectrum_centre1}
\end{align}
These configurations aim to uniformly distribute the spectra of \acp{led} and the passband of optical filters within the visible light spectrum.

\subsubsection{Configuration with \ac{bads} algorithm}
The parameter searching using the \ac{bads} algorithm with \ac{wd} strategy is similar to that in Section~\ref{subsubsec:SD_optimised} except that the black-box optimisation problem is redefined as follows:
\begin{align}
& \underset{\mathcal{X}}{\text{maximise}} &&\bar{C}^{\rm WD}(\mathcal{X})~\text{with}~\mathcal{X}=\left\{\lambda_{n_{\rm t}}^{\rm led,c},\lambda_{n_{\rm r}}^{\rm of,c},\Delta\lambda^{\rm of}\right\} \\
&\text{subject to} &&\lambda_{\rm min}\leq\lambda_{n_{\rm t}}^{\rm led,c}\leq\lambda_{\rm max}~\text{for}~n_{\rm t}=1,\cdots,I, \\
& ~ && \lambda_{\rm min}\leq\lambda_{n_{\rm r}}^{\rm of,c}\leq\lambda_{\rm max}~\text{for}~n_{\rm r}=1,\cdots,I, \\
& ~ && \Delta\lambda^{\rm of} > 0.
\end{align}

{\color{blue} Fig.~\ref{fig:SE_WDM} shows the average achievable rates achieved by joint multiplexing systems with \ac{wd} strategy against the number of elements. The remaining system parameters are the same as those listed in Table~\ref{table:Parameters}. It can be observed that the highest average achievable rate by the systems without \ac{mimo} processing is less than 400~Mbps at two elements. This demonstrates the severe performance penalty caused by the bandpass shift issue and the excessive inter-colour interference with a large number of \acp{wd}. Therefore, the \ac{mimo} processing is important for joint multiplexing systems with \ac{wd} strategy. In contrast, the average rates achieved by systems with \ac{mimo} processing increase with the number of elements consistently when the number of elements is small. However, when the number of elements is between 7 to 9, a further increase in the number of elements no longer provides an effective boost to the average achievable rate. In actuality, the achievable rates drop with an increase of the number of elements in the cases of empirical configurations. In the cases of configuration with a \ac{bads} algorithm, the performance degradation with an optical element increase can be avoided, but no further improvement can be observed either, as shown in Fig.~\ref{fig:SE_WDM}. This is because the use of optical filters with a narrow passband actively block a significant amount of optical power from \acp{led}. The narrower the optical filter passband, the more power loss occurs, which leads to a more severe degradation in the overall \acp{snr} and achievable rate. In addition, the increased number of elements leads to a more severe spectrum overlap between adjacent \ac{wd} channels, which causes significant correlations between \ac{wd} channels. Consequently, the multiplexing channels with smaller eigenvalues are severely degraded. Compared to cases with \ac{sd} strategy, all \acp{led} and \acp{pd} have the same orientation and positions. When there is a good alignment, all multiplexing channels are achieving high \ac{snr} and capacity. When there is a bad alignment, none of the channels show a good performance. The issue is more severe in the case of random orientation scenarios. This situation is quite different from the \ac{sd} systems where the chance of at least one or a few good channels is much higher due to the better spatial diversity. Note that the performance gap between upward scenario and corresponding random orientation scenario is smaller in \ac{sd} systems.
}


\subsection{Spatial clustering with wavelength division strategy}
\label{subsec:HMP}
In Section~\ref{subsec:SMP} and \ref{subsec:WDM}, the limitation of \ac{sd} and \ac{wd} strategies has been demonstrated. In this section, the configurations based on the approach of `division in either the space or wavelength domain' are considered. The proposed strategy is named as \ac{scwd}. The basic idea of the \ac{scwd} scenario is to divide the \acp{led} and \acp{pd} into multiple groups. The optical elements in each group are clustered with the same positions and orientations, as illustrated in Fig.~\ref{fig:configuration_scenarios}~(d). Thus, the optical elements in different groups/clusters have different spatial features to decorrelate the channel. On the other hand, \acp{led} in the same cluster have different spectra, and optical filters with different passbands are mounted on \acp{pd} in the same cluster. This wavelength division can remove the channel correlation between  optical elements within the same group/cluster.

{\color{blue} 
Next, the details about the proposed \ac{scwd} strategy will be introduced. Assume that the \ac{mimo}-\ac{ofdm} system has $I$ \acp{led} and \acp{pd} divided into $L$ clusters, where $1\leq L\leq I$. The configuration of spatial variables is similar to that introduced in Section~\ref{subsec:SMP} except that the configuration is with respect to each spatial cluster instead of each individual optical element. On the transmitter side, the $l$th \ac{led} cluster has a position of:
\begin{align}
\mathbf{p}_{{\rm c},l}^{\rm led}=[\mathtt{x}_{l},\mathtt{y}_{l},\mathtt{z}_{\rm tx}],
\end{align}
for $l=1,2,\cdots,L$, where $\mathtt{x}_{l}$ and $\mathtt{y}_{l}$ are the horizontal coordinates of the $l$th \ac{led} cluster, which are configurable parameters. All \ac{led} clusters have the same orientation of $[0,0,-1]$. In the case of an upward facing receiver, the $l$th \ac{pd} cluster has an orientation vector of:
\begin{align}
\mathbf{o}_{{\rm c},l}^{\rm pd,\uparrow}=\left[\cos\omega_{l}\sin\theta_{l},\sin\omega_{l}\sin\theta_{l},\cos\theta_{l}\right],
\end{align}
where $\omega_{l}$ and $\theta_{l}$ are the azimuth angle and elevation angle of the $l$th \ac{pd} cluster, which are defined based on the characteristics of \ac{pr} and \ac{hr} using \eqref{eq:pr_angle}, \eqref{eq:hr_angle1} and \eqref{eq:hr_angle2}. Similar to the \ac{sd} strategy, $\phi_{1/2}$, $\mathfrak{m}_{\rm fov}$ and $(\theta_{\rm pd})$ are also spatial domain configurable parameters. The number of optical elements in the $l$th cluster is defined as:
\begin{algorithm}[t]
	\SetAlgoLined
	\fontsize{10pt}{10pt}\selectfont
	Initialise cluster index: $l=1$ \\
	\For{$i=1,2,\cdots,I$}{
		$M_{\rm sum}=\sum_{\hat{l}=1}^{l}M_{\hat{l}}$ \\
		\If{$i>M_{\rm sum}$}{Cluster index increment: $l=l+1$	
		}
		$m=i-M_{\rm sum}+M_{l}$ \\
		Variable mapping: $\mathtt{x}_{i}=\mathtt{x}_{l}$, $\mathtt{y}_{i}=\mathtt{y}_{l}$, $\omega_{i}=\omega_{l}$, $\theta_{i}=\theta_{l}$, $\lambda^{\rm led}_{i}=\lambda^{\rm led}_{{\rm c},m}$, $\lambda^{\rm of}_{i}=\lambda^{\rm of}_{{\rm c},m}$ \\
	}
	\caption{Cluster-to-element variable mapping}
	\label{algo:spatial_cluster}	
\end{algorithm}
\begin{align}
M_{l}&=\left\{\begin{array}{lr} \lceil I/L\rceil &: l\leq \mod(I,L) \\
\lfloor I/L\rfloor &: \text{otherwise}
\end{array}
\right.,
\end{align}
which ensure the sizes of different clusters are similar to a maximum difference of one element. Note that $\lceil\cdot\rceil$ and $\lfloor\cdot\rfloor$ are defined as the ceiling and floor operators, respectively. This also implies that the maximum required number of \ac{wd} is $\lceil I/L\rceil$. Within the $l$th cluster, the central wavelength of the $m$th \ac{led} is defined as $\lambda^{\rm led}_{{\rm c},m}$ on the transmitter side for $m=1,2,\cdots,M_{l}$. On the receiver side, the passband centre of the $m$th optical filter is defined as $\lambda^{\rm of}_{{\rm c},m}$. Opposite to the spatial parameter configurations, the wavelength domain parameters are identical in different clusters. Similar to the \ac{wd} strategy, in addition to $\lambda^{\rm led}_{{\rm c},m}$ and $\lambda^{\rm of}_{{\rm c},m}$, the optical filter passband with $\Delta\lambda^{\rm of}$ is also a configurable parameter. By unifying some of the parameters, the system configuration process can be significantly simplified. If the configuration is with respect to each element, the number of configurable parameters will be $4I+4$. With the cluster-based configuration, there are $2(L+\lceil I/L\rceil)+4$ configurable parameters, which include: $\mathtt{x}_{l}$, $\mathtt{y}_{l}$,  $\lambda_{{\rm c},m}^{\rm led}$, $\lambda_{{\rm c},m}^{\rm of}$, $\phi_{1/2}$, $\mathfrak{m}_{\rm fov}$, $(\theta_{\rm pd})$, $\Delta\lambda^{\rm of}$.}

With cluster-based parameters $\mathtt{x}_{l}$, $\mathtt{y}_{l}$, $\omega_{l}$, $\theta_{l}$, $\lambda^{\rm led}_{{\rm c},m}$, $\lambda^{\rm of}_{{\rm c},m}$, the corresponding parameters with respect to the $i$th optical element $\mathtt{x}_{i}$, $\mathtt{y}_{i}$, $\omega_{i}$, $\theta_{i}$, $\lambda^{\rm led}_{i}$, $\lambda^{\rm of}_{i}$ can be obtained using Algorithm~\ref{algo:spatial_cluster}. With a given set of spatial and wavelength domain parameters, the average achievable rate with a specific number of clusters $L$ can be calculated. However, it is not straightforward to select $L$ that achieves the highest average achievable rate. Therefore, the average achievable rates with all possible numbers of clusters $\bar{C}_{L}^{\rm SCWD}$ are evaluated for $L=1,2,\cdots,I$. Finally, we find the result achieves the highest average achievable rate among the evaluations with different $L$ as:
\begin{align}
 \bar{C}^{\rm SCWD}=\mathrm{max}~\left\{\bar{C}_{L=1}^{\rm SCWD},\bar{C}_{L=2}^{\rm SCWD},\cdots,\bar{C}_{L=I}^{\rm SCWD}\right\}.
 \label{eq:average_capacity_scwd}
\end{align}

%

\subsubsection{Empirical configuration}
Regarding the empirical configuration, some spatial domain parameters are the same as those used in the \ac{sd} strategy for a fairer comparison: $\phi_{1/2}=60\degree$, $\mathfrak{m}_{\rm fov}=1.4738$, $\theta_{\rm pd}=40\degree$. The values of $x_{l}$ and $y_{l}$ with different number of clusters $L$ are selected based on Fig.~\ref{fig:LED_layout}. The configuration of $\lambda^{\rm led}_{{\rm c},m}$, $\lambda^{\rm of}_{{\rm c},m}$ and $\Delta\lambda^{\rm of}$ are similar to \eqref{eq:passband_width1} and \eqref{eq:spectrum_centre1} in the \ac{wd} strategy empirical configuration:
\begin{align}
\Delta\lambda^{\rm of}&=(\lambda_{\rm max}-\lambda_{\rm min})/\lceil I/L\rceil,
\label{eq:passband_width2} \\
\lambda_{{\rm c},m}^{\rm led}&=\lambda_{{\rm c},m}^{\rm of}=\lambda_{\rm min}+(m-0.5)\Delta\lambda^{\rm of}.
\label{eq:spectrum_centre2}
\end{align}
After evaluation of $\bar{C}_{L}^{\rm SCWD}$ for all possible $L$, the highest achievable rate $\bar{C}^{\rm SCWD}$ is selected using \eqref{eq:average_capacity_scwd}.


\subsubsection{Configuration with \ac{bads} algorithm}
The black-box optimisation problem for \ac{scwd} strategy is defined as:
\begin{align}
& \underset{\mathcal{X}}{\text{maximise}} &&\bar{C}_{\rm L}^{\rm SCWD}(\mathcal{X})~\text{with}~\mathcal{X} =\left\{\mathtt{x}_{l}, \mathtt{y}_{l},  \lambda_{{\rm c},m}^{\rm led}, \lambda_{{\rm c},m}^{\rm of}, \phi_{1/2}, \mathfrak{m}_{\rm fov}, (\theta_{\rm pd}), \Delta\lambda^{\rm of}\right\} \\
&\text{subject to} &&0<\frac{\mathtt{x}_{l}}{W_{\rm room}}<1~\text{for}~l=1,2,\cdots,L, \label{eq:con1} \\
& ~ && 0<\frac{\mathtt{y}_{l}}{L_{\rm room}}<1~\text{for}~l=1,2,\cdots,L, \\
& ~ && \lambda_{\rm min}\leq\lambda_{{\rm c},m}^{\rm led}\leq\lambda_{\rm max}~\text{for}~m=1,2,\cdots,M_{l}, \\
& ~ && \lambda_{\rm min}\leq\lambda_{{\rm c},m}^{\rm of}\leq\lambda_{\rm max}~\text{for}~m=1,2,\cdots,M_l, \label{eq:con4} \\
& ~ && 0\degree<\phi_{1/2}\leq 60\degree, \\
& ~ && m_{\rm fov}\geq 1, \\
& ~ && (0\degree\leq\theta_{\rm pd}\leq 90\degree) \\
& ~ && \Delta\lambda^{\rm of} > 0,
\end{align}
Note that solving this problem only finds the highest average achievable rate  with $L$ clusters. Therefore, the \ac{bads} algorithm need to be executed for $I$ times for each initial solution so that the achievable rates for all possible $L$ are obtained. It is challenging to include the parameter $L$ as one of the optimisation variables, as the number of parameters in the system configuration is a function of $L$.

\begin{figure}[!t]
	\centering
    \begin{subfigure}{.5\textwidth}
	    \centering
	    \includegraphics[width=1\linewidth]{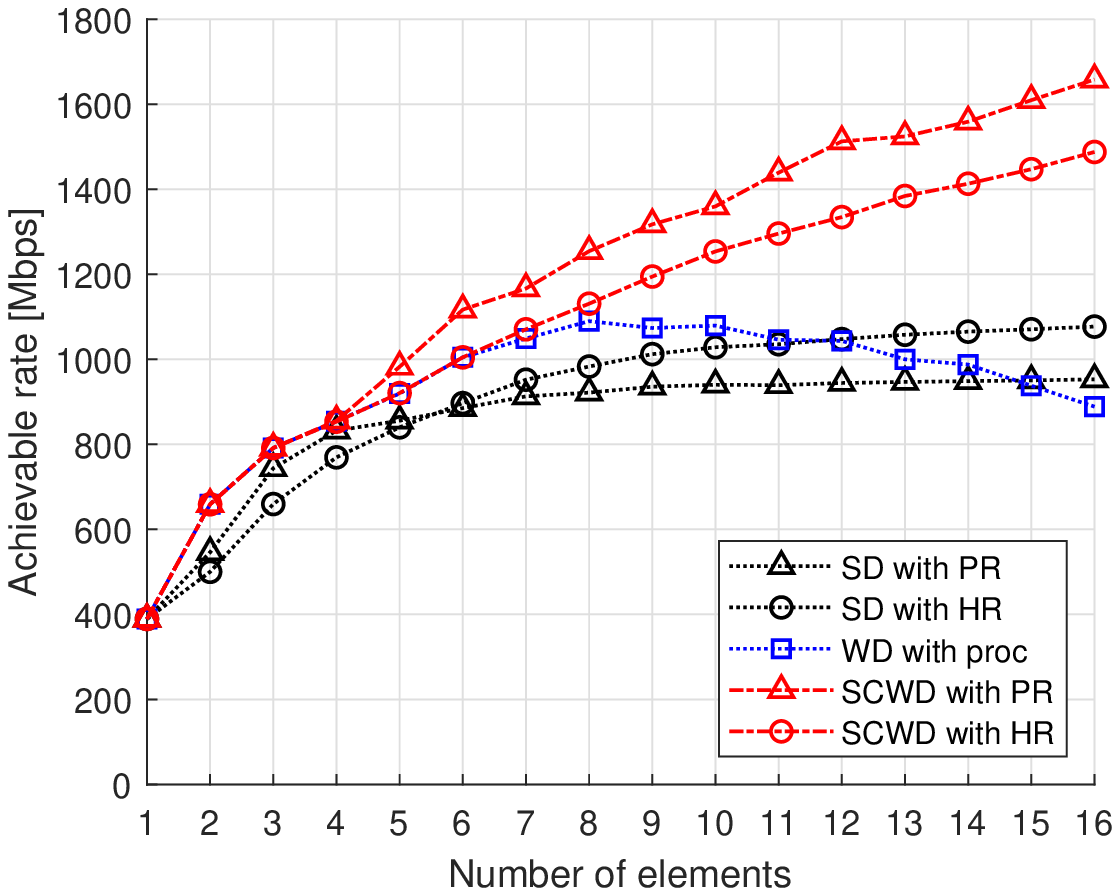}
    	\caption{}
    	\label{fig:HMP_emp_upward}
    \end{subfigure}%
    \begin{subfigure}{.5\textwidth}
	    \centering
	    \includegraphics[width=1\linewidth]{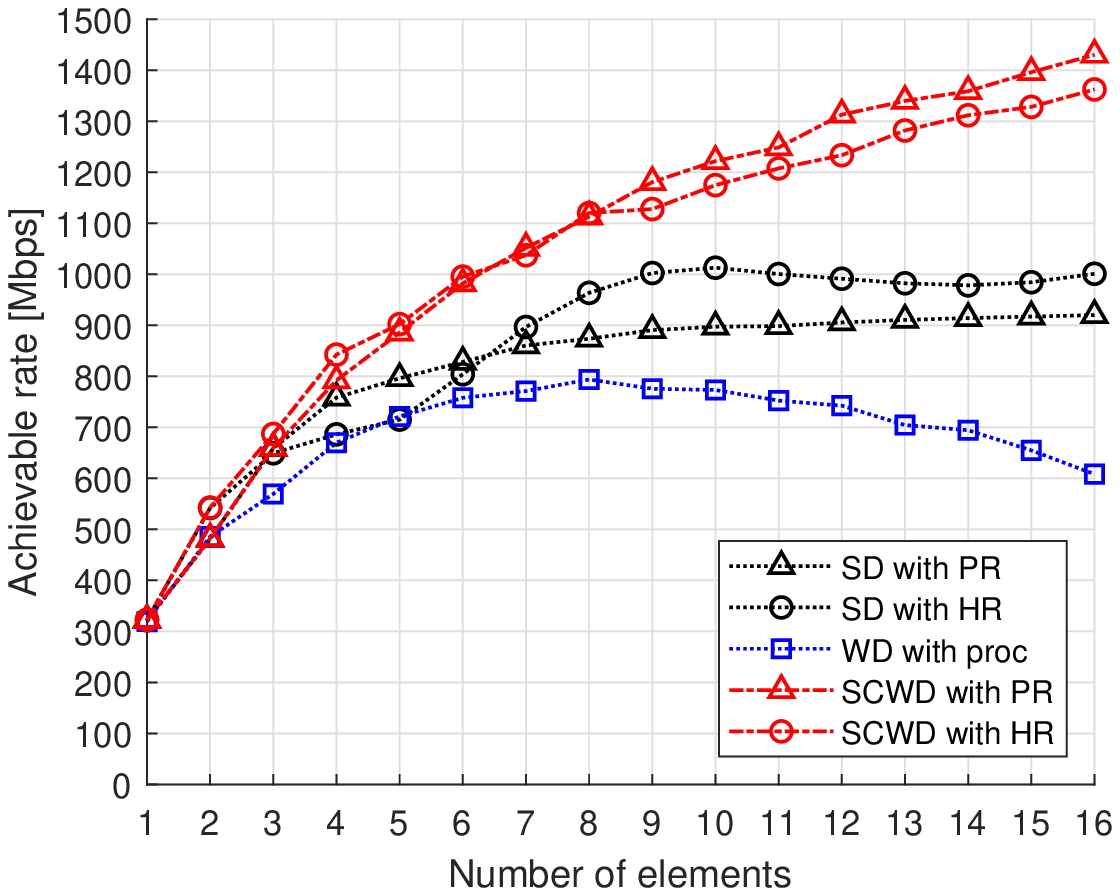}
    	\caption{}
    	\label{fig:HMP_emp_random}
    \end{subfigure}
    \\
    \begin{subfigure}{.5\textwidth}
	    \centering
    	\includegraphics[width=1\linewidth]{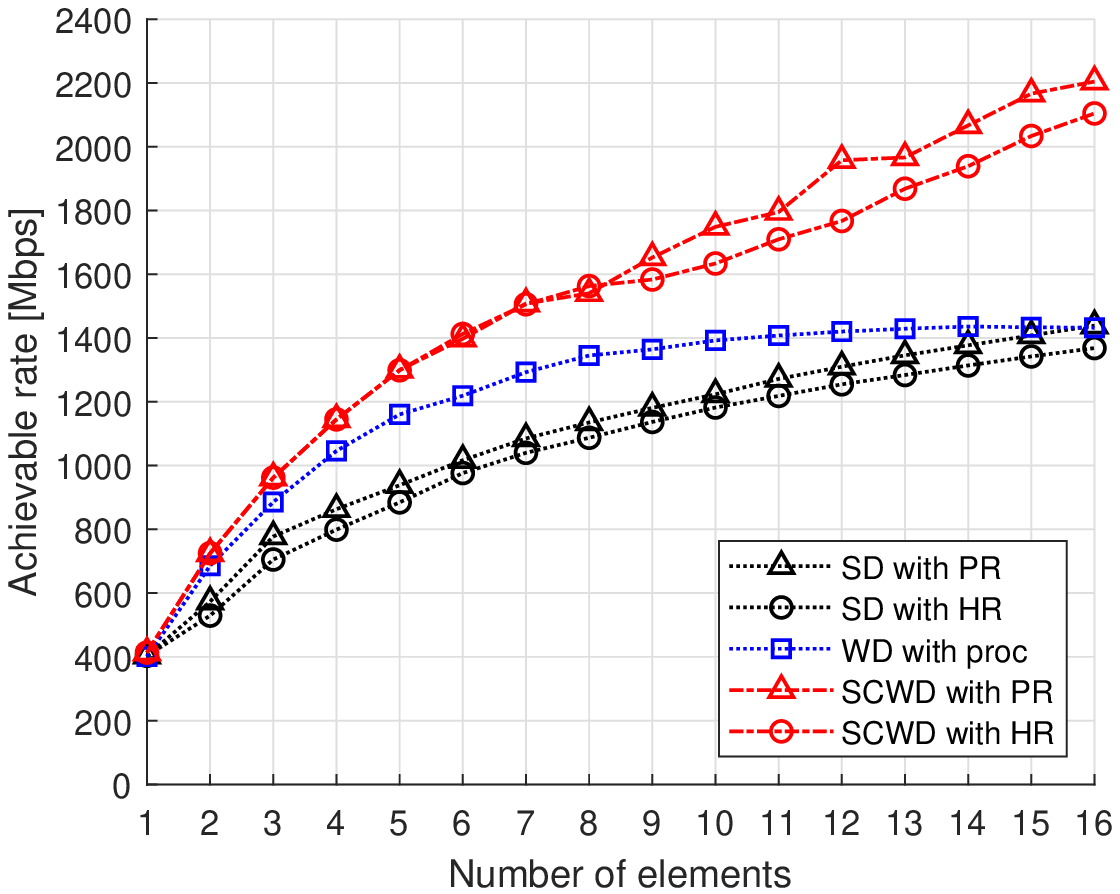}
    	\caption{}
	    \label{fig:HMP_opt_upward}
    \end{subfigure}%
    \begin{subfigure}{.5\textwidth}
	    \centering
    	\includegraphics[width=1\linewidth]{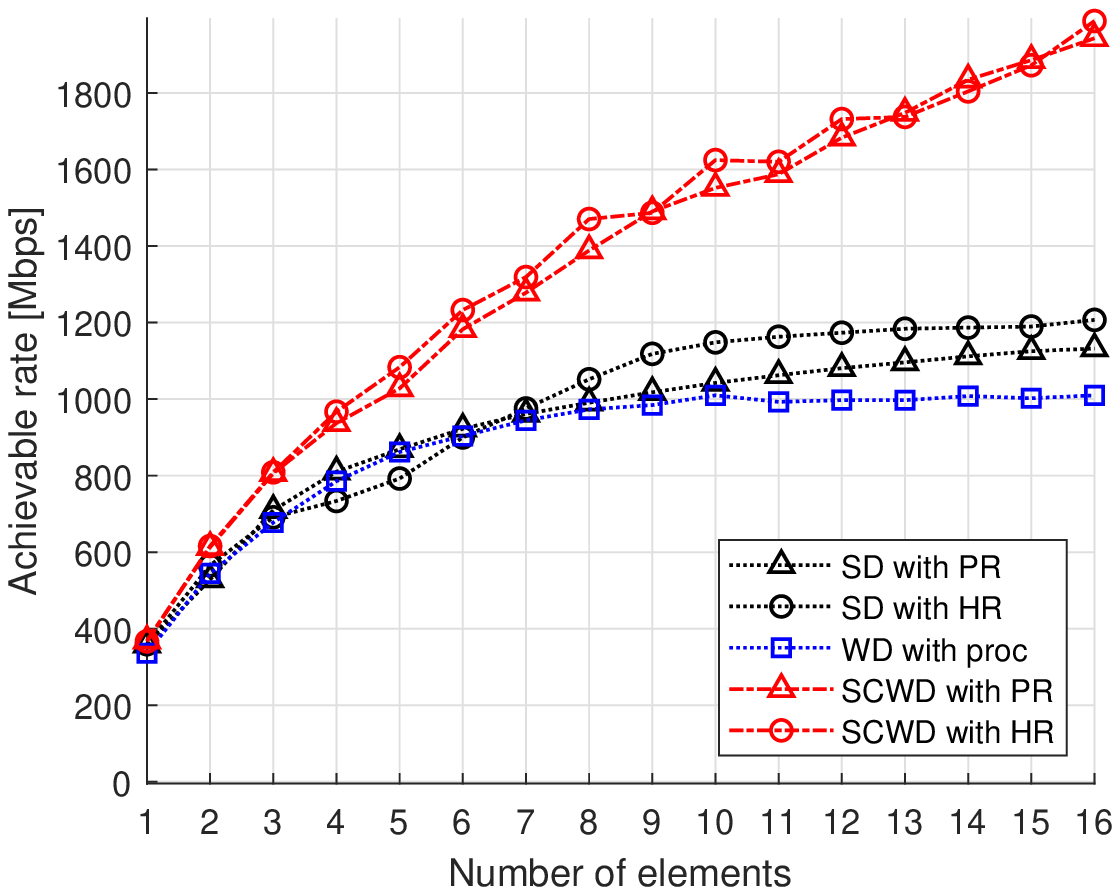}
    	\caption{}
	    \label{fig:HMP_opt_random}
    \end{subfigure}    
	\caption{Average achievable rate of joint multiplexing systems with \ac{scwd} strategy against number of elements. (a) Empirical configurations with upward receiver orientation. (b) Empirical configurations with random receiver orientation. (c) BADS-based configurations with upward receiver orientation. (d) BADS-based configurations with random receiver orientation.}
	\label{fig:HMP_SE}
	\vspace{-7mm}
\end{figure}

Fig.~\ref{fig:HMP_SE} shows the average rates achieved by the joint multiplexing systems with the \ac{scwd} strategy against different numbers of elements. In addition, the results with \ac{sd} and \ac{wd} strategies are also included for comparison. The remaining parameters are listed in Table~\ref{table:Parameters} if they are not specified. The four plots show cases with different receiver orientation scenarios and with different types of configurations, respectively. It can be observed that with an increase in the number of elements, the average achievable rate by the \ac{scwd} strategy systems also increases consistently, even when the number of elements is greater than 10. With 16 optical elements, the data rates achieved by \ac{scwd} systems are in the range of 1300 to 2200 Mbps, while those achieved by either \ac{sd} or \ac{wd} systems fall in the range of 500 to 1400 Mbps. The achievable rate improvement is between 36\% and 74\% compared to the \ac{sd} strategy, and the improvement is between 47\% and 135\% compared to the \ac{wd} strategy. In general, the \ac{scwd} systems with \ac{pr} shows a slightly better performance at a high number of elements compared to those with \ac{hr}.

\subsection{Insights into the system parameter configurations and \ac{scwd} strategy}
\label{subsec:insights}
In this subsection, we will discuss the difference between the parameters used in the empirical configurations and configurations with the \ac{bads} algorithm. We will also explain why the \ac{scwd} strategy is able to provide extra multiplexing gains when compared to \ac{sd} and \ac{wd} strategies. For simplicity, the cases with a \ac{pr} and the upward receiver orientation are demonstrated as an example. 
\begin{figure}[t!]
	\centering
	\begin{subfigure}{.51\textwidth}
		\centering
		\includegraphics[width=1\linewidth]{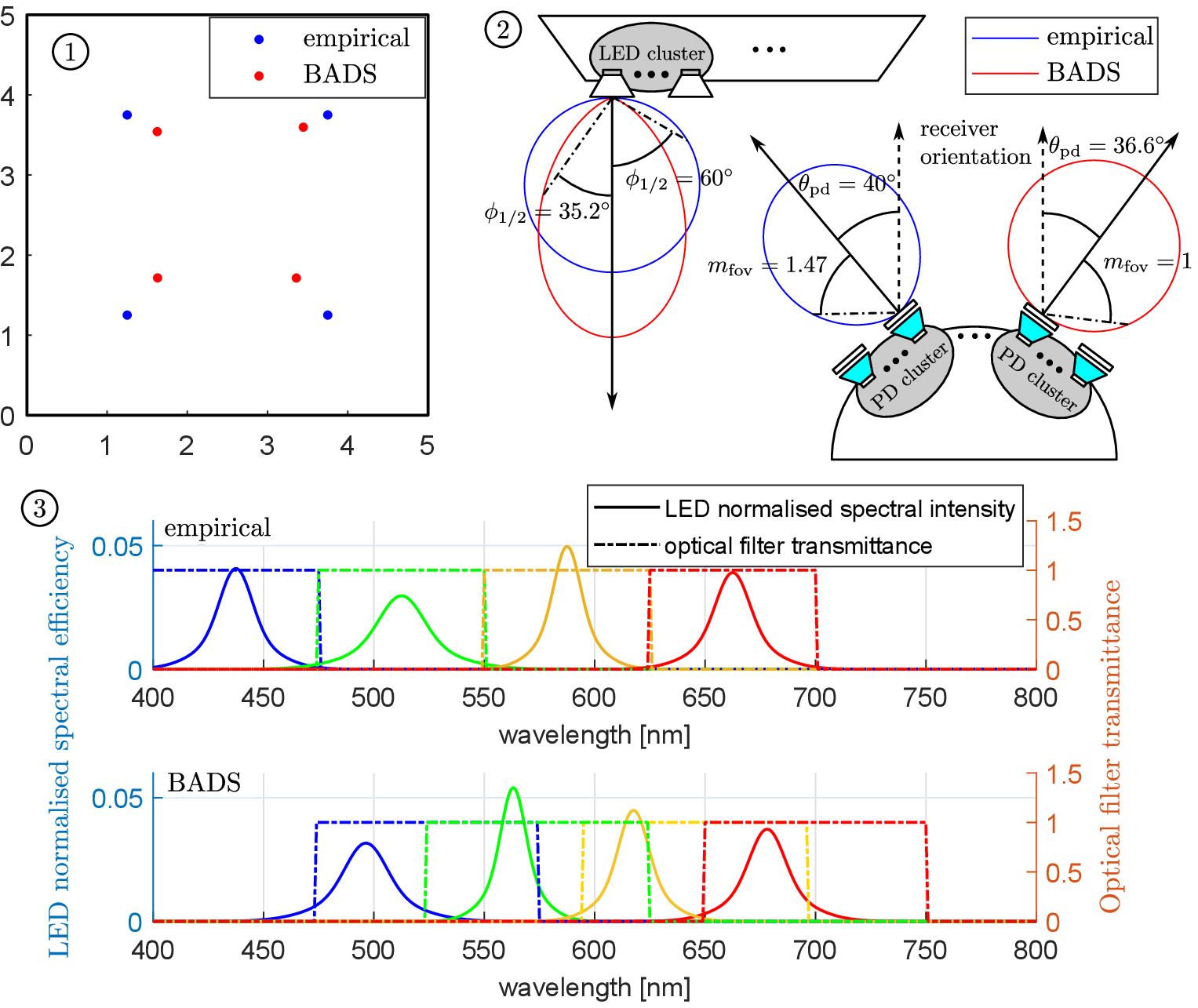}
		\caption{}
		\label{fig:configuration_compare}
	\end{subfigure}%
	\begin{subfigure}{.49\textwidth}
		\centering
		\includegraphics[width=1\linewidth]{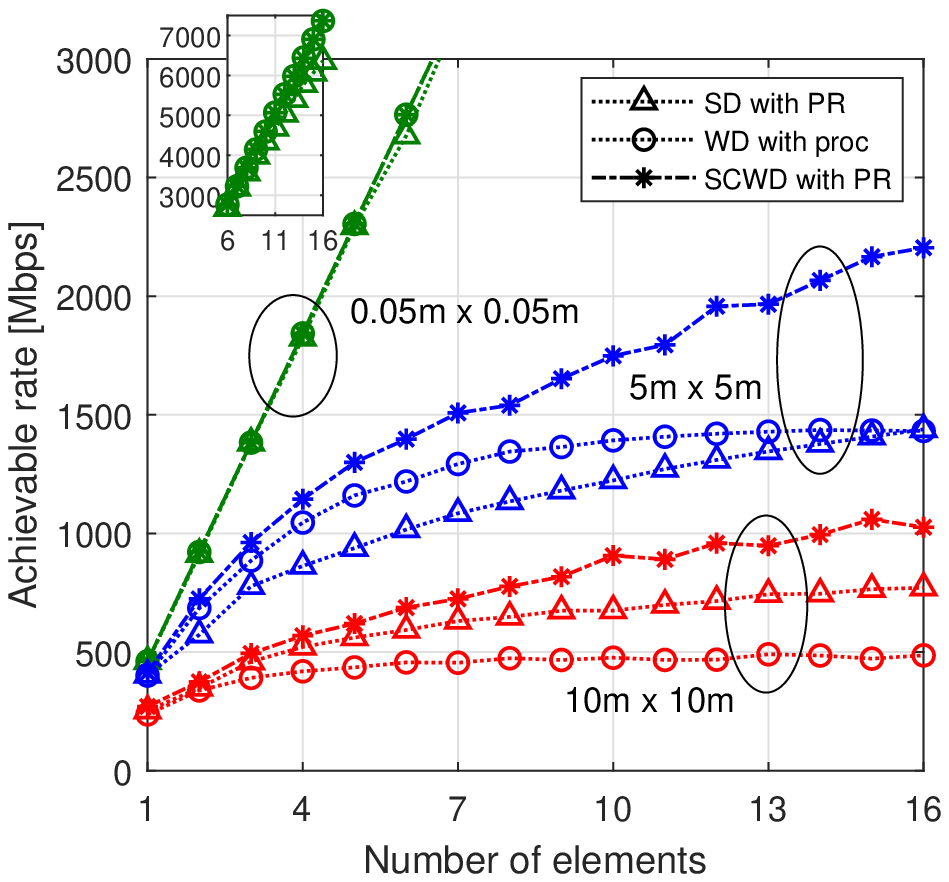}
		\caption{}
		\label{fig:case_studies}
	\end{subfigure}
	\caption{(a) Comparison between empirical and \ac{bads}-based configuration parameters used in 16-elements joint multiplexing systems with SCWD strategy. The demonstrated systems use \ac{pr} and upward receiver orientation scenario. 1) LED  cluster position layout. 2) Angular parameters. 3) Wavelength-dependent parameters. (b) Achievable rate of systems using optimised configurations with \ac{pr} and upward receiver orientation scenario against number of elements in two special conditions. Condition~1: larger room of size $\rm 10~m\times10~m$. Condition~2: extremely small space of size $\rm 0.05~m\times0.05~m$. Default condition: room size of $\rm 5~m\times 5~m$.}
	\label{}
	\vspace{-7mm}
\end{figure}
The comparison between parameters used in empirical and \ac{bads}-based configurations with 16 elements is shown in Fig.~\ref{fig:configuration_compare}. In both configurations, there are four spatial clusters, as demonstrated in Fig.~\ref{fig:configuration_compare}~1). Regarding the geometric parameters, the spatial positions of \ac{led} clusters are closer to the room centre in the case of \ac{bads}-based configuration compared to the empirical configuration. In addition, a smaller half-power semiangle of $\phi_{1/2}=35.2\degree$ is used on the transmitter side and a smaller \ac{fov} coefficient of $m_{\rm fov}=1$ is used on the receiver side compared to the empirical configuration, as shown in Fig.~\ref{fig:configuration_compare}~2). These parameter changes lead to weaker spatial channel correlation and thereby achieving higher overall multiplexing gain. Regarding the wavelength domain parameters, four \acp{wd} are used within each spatial cluster, as shown in Fig.~\ref{fig:configuration_compare}~3). Compared to the case of empirical configuration, all \ac{led} spectra shift to the longer wavelengths with smaller separations in the case of \ac{bads}-based configuration. This is because the \ac{pd} has a higher responsivity at a longer wavelength and the additional inter-colour interference introduced by smaller separation causes negligible distortion. In addition, the passband of optical filters are extended to longer wavelengths in the case of \ac{bads}-based configuration, which compensate bandpass shift issue when light incident angle is large.

In order to demonstrate the characteristics of the \ac{scwd} strategy, we can evaluate the performance of joint multiplexing systems using \ac{bads}-based configurations with \ac{pr} and upward receiver orientation scenario under several conditions. In addition to cases with normal conditions (i.e. a room size of $\rm 5~m \times 5~m \times 3~m$) shown in Fig.~\ref{fig:HMP_opt_upward}, we further evaluate two special conditions. In the first condition, we show the transmissions in a large room of size $\rm 10~m \times 10~m \times 3~m$. The remaining system parameters are identical to those used in Sections~\ref{subsec:SMP}, \ref{subsec:WDM} and \ref{subsec:HMP}. Due to the increase in coverage area, the average link distance becomes greater and the misalignment issue due to random user position and orientation is more severe. These factors lead to a further decrease in the data rate achieved by joint multiplexing systems, as shown in Fig.~\ref{fig:case_studies}. The achievable rate decrease is more severe for the case of \ac{wd} strategy, where random users are more likely to receive a weaker signal as all \acp{led} are located in the room centre. Therefore, the achievable rate of the \ac{wd} strategy is considerably lower than the case of \ac{sd} strategy with the same number of elements. Due to the inferior performance of system with \ac{wd} strategy, the \ac{scwd} strategy provides a decreased achievable rate improvement against its counterparts compared to the case with $\rm 5~m \times 5~m\times 3~m$ room. In the second special condition, we show that transmissions in an extremely small indoor space of size $\rm 0.05~m \times 0.05~m \times 3~m$. With such a small coverage area and the upward receiver scenario, all users experience a minimised link distance with good alignment. Consequently, the passband shift issue in systems with the \ac{wd} strategy can be avoided. Note that a half-power semiangle of $\phi_{1/2}=1.5\degree$ is also used in the case of \ac{wd} strategy to maximise the detected optical power. In the case of \ac{sd} strategy, performance degradation due to greater link distance and misalignment is avoided. In addition, a \ac{bads}-based configuration solution with a very collimated light beam and light reception pattern is obtained in case of the \ac{sd} strategy. Due to all the above factors, the joint multiplexing systems with \ac{wd} and \ac{sd} strategies can operate in ideal conditions. Specifically, the multiplexing gain scales almost linearly with the number of elements, as shown in Fig.~\ref{fig:case_studies}. In this condition, despite the minor performance gap between \ac{wd} and \ac{sd} systems, the multiplexing gain of both systems are approaching the theoretical limit of multiplexing systems. Consequently, the system with \ac{scwd} strategy exhibits a similar ideal performance. 


Despite the undesired performance exhibited in the two special conditions, using a \ac{mimo} \ac{vlc} system to cover an extremely large or extremely small area is unlikely in practice. As long as the following two conditions are fulfilled, \ac{scwd} strategy will provide a considerable improvement: 1) there is a minor performance gap between joint multiplexing systems with \ac{sd} and \ac{wd} strategies, 2) multiplexing transmissions in individual spatial and wavelength domains are inefficient. In general conditions of where a medium sized room is covered, similar achievable rates are likely to be achieved by systems with \ac{sd} and \ac{wd} strategies. In addition, the efficiency of \ac{vlc} multiplexing systems is limited by the random user position/orientation and channel correlation. Therefore, the \ac{scwd} strategy is likely to provide a remarkable achievable rate improvement, as shown in Fig.~\ref{fig:HMP_SE}. The configuration problem in joint multiplexing systems is similar to a power allocation problem in a multi-carrier transmission system with two channels. If both channels have similar qualities, loading them with power leads to a higher performance compared to loading all the power to one channel. However, if one of the channels is unreliable or the spectral efficiency scales linearly with the power increase, loading two channels will no longer be beneficial.

{\color{blue}
\subsection{Practical implementation and illumination}
For implementations of the \ac{vlc} \ac{mimo} joint multiplexing system with the proposed \ac{scwd} strategy, there are several practical solutions. For example, it is challenging to mount more than ten \acp{pd} on the same side of a portable device, while it is easier to mount a cluster of a few \acp{pd} on each side of a device with a different orientation \cite{9120825}. For small devices, such as smart phones, mounting one or two \acp{pd} on each side of the device is realistic. The system with a greater number of \acp{pd} can be implemented on larger devices such as tablets, laptop or \ac{iot} device. On the transmitter side, the \ac{scwd} strategy gather multiple \acp{led} with different wavelengths in the same cluster. This feature provides an opportunity to generate white light illumination by manipulating the intensity ratio of \acp{led} with different colours \cite{7114351}. In the cases where the \acp{led} in the cluster are insufficient to provide the desired white light, additional \acp{led} with suitable wavelengths can be installed in the \ac{led} cluster so that the combined light reaches the illumination requirement.
}

\section{Conclusions}
\label{sec:conclusion}
A framework for a \ac{vlc} \ac{mimo}-\ac{ofdm} joint multiplexing system with a channel dependent on spatial, wavelength and frequency domain characteristics is proposed. This study focused on investigating the suitable system configurations for the \ac{vlc} joint multiplexing system with different strategies in order to achieve a higher multiplexing gain, thereby achieving a higher aggregate transmission data rate. The proposed \ac{scwd} strategy shows a considerable improvement in the average achievable rate compared to cases with \ac{sd} and \ac{wd} strategies. The use of the \ac{bads} black-box optimisation algorithm improves the achievable rates further. The simulation results demonstrate an average achievable rate improvement of 36\% to 135\% with 16 \acp{led}/\acp{pd} by using the proposed \ac{scwd} strategy compared to those achieved with \ac{sd} and \ac{wd} strategies. This work also demonstrated the possibility of using a \ac{vlc} \ac{mimo}-\ac{ofdm} system with limited modulation bandwidth to achieve multi-Gbps transmission data rates, which shows the potential of \ac{vlc}/\ac{lifi} systems in next generation wireless networks in terms of achievable rate capability. {\color{blue} In future studies, the scenarios with multiple access or multi-user \ac{mimo} will be investigated.}



%

\ifCLASSOPTIONcaptionsoff
  \newpage
\fi

\bibliographystyle{IEEEtran}
\bibliography{MIMO_OFDM_joint_multiplexing}




\end{document}